\documentclass[12pt]{article}

\usepackage{epsfig}
\usepackage{amssymb}
\usepackage{amsfonts}

\usepackage{color}

\def\be{\begin{equation}}
\def\ee{\end{equation}}
\def\ba{\begin{array}{c}}
\def\ea{\end{array}}

\newcommand{\bea}{\begin{eqnarray}}
\newcommand{\eea}{\end{eqnarray}}

\begin{document}

\titlepage

%\vspace{.35cm}

 \begin{center}{\Large \bf

Few-grid-point simulations of Big Bang singularity in quantum cosmology

  }\end{center}

 \begin{center}

\vspace{8mm}

  {\bf Miloslav Znojil} $^{1,2}$

\end{center}

\vspace{8mm}

  $^{1}$
 {The Czech Academy of Sciences,
 Nuclear Physics Institute,
 Hlavn\'{\i} 130,
250 68 \v{R}e\v{z}, Czech Republic, {e-mail: znojil@ujf.cas.cz}}

 %\footnote{{e-mail: znojil@ujf.cas.cz}

 $^{2}$
 {Department of Physics, Faculty of
Science, University of Hradec Kr\'{a}lov\'{e}, Rokitansk\'{e}ho 62,
50003 Hradec Kr\'{a}lov\'{e},
 Czech Republic}

%\newpage

\section*{Abstract}

In the context of the current
lack of compatibility of the
classical and quantum
approaches to gravity, exactly
solvable elementary pseudo-Hermitian
quantum models are analyzed
supporting the
acceptability of a point-like form of
Big Bang.
The purpose is served by
a hypothetical (non-covariant) identification
of the ``time of Big Bang''
with the Kato's exceptional-point parameter $t=0$.
Consequences
(including the
ambiguity of the
patterns of unfolding of the singularity after Big Bang)
are studied in detail. In particular,
singular values of
the observables are shown useful
in the analysis.

%\newpage

\subsection*{Keywords}.

unfolding Big Bang in classical physics;

spatial-grid-point-position as quantum observable;

quantized Big Bang as Kato's exceptional point;

non-Hermitian toy models;

bracketing property of singular values;

\newpage

\section{Introduction}

\begin{figure}[h]                     %instead of \begin{figure}[t]
\begin{center}                         %instead of \begin{center}
\epsfig{file=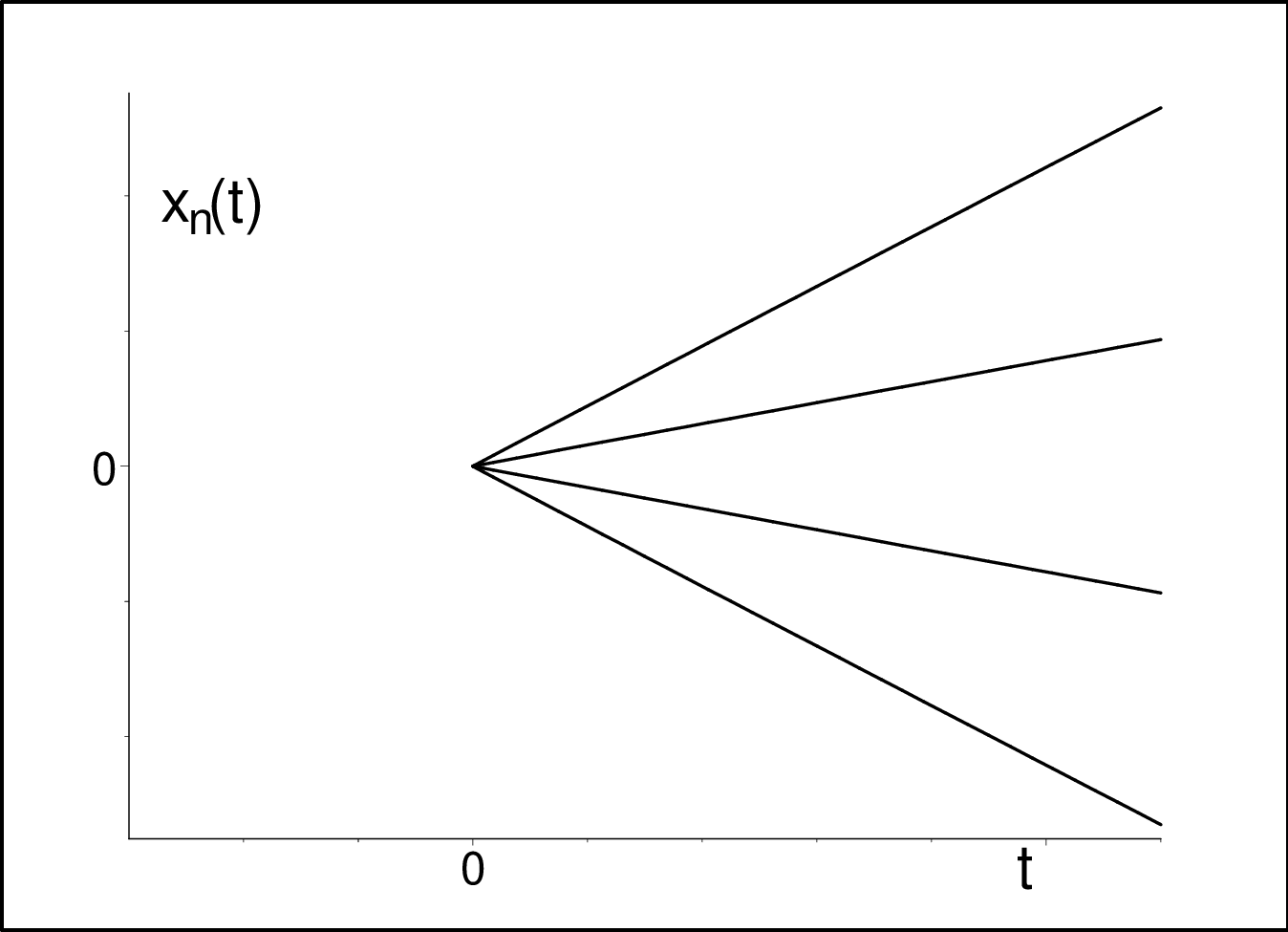,angle=270,width=0.6\textwidth}
\end{center}                         %instead of \end{center}
\vspace{-2mm}\caption{The ``Big Bang'' birth of the Universe, with the
expansion of the space sampled by four grid points $x_n(t)$
(non-covariant classical-physics picture, arbitrary units).
%
%The sceptic's ``no-prehistory''
%alternative to Figure~\ref{2pic}.
}
\label{3pic}
\end{figure}

The determination of the age of our Universe (i.e., of its estimated
$13.787\pm 0.02$ billion years) belongs, undoubtedly, among the most
impressive achievements of experimental astronomy
\cite{Planck,Planckb}. In parallel, the hypothetical birth of the
Universe at zero diameter and zero time $t=0$ (called Big-Bang
singularity, cf. its schematic picture in Figure~\ref{3pic}) finds a
consistent explanation and interpretation in the Einstein's
formalism of general relativity.

%Big Bang 1 (linear):
%
%plot({0.1*sqrt(0+0.1*s^1.99),-0.1*sqrt(0+0.1*s^1.99),0.2*sqrt(0+0.2*s^1.99),
% -0.2*sqrt(0+0.2*s^1.99)},s=-6..12,color=black,axes=framed,tickmarks=[2,2]);
%
%%
%
%Big Bang unfolding  (non-Hermitian matrix,
%linear-evolution
%sample, arbitrary units,
%spectrum = real at $s \leq 0$,
%exceptional point of the fourth order, EP4).

In the latter framework, the question of
``what happened before Big Bang?''
is usually considered irrelevant.
Incidentally, a different opinion
has been formulated by
Penrose \textcolor{black}{who conjectured that}
%resurrected
%the old (hinduistic?) cyclic-cosmology hypothesis by which
the Universe might have existed even before the Big Bang~\cite{Penrose}.
Its prehistory is to be perceived as eternal,
composed of a series of
\textcolor{black}{Aeons}, well-separated
by their collapses
\textcolor{black}{(or, more precisely,
by the stages of an unlimited
expansion tractable also as an ``entropic death''),}
followed by a
%Big-Bang-like
rebirth
\textcolor{black}{and subsequent
expansion
(notice that the term
``Big Bang'' itself has originally been
coined as a mockery (!)).}

%
%(cf. a schematic sample of such a rebirth in Figure \ref{2pic}).

%
%EP-like unavoided crossing:
%at $s = 0$ (non-Hermitian matrices, schematic
%illustration, arbitrary units, exceptional point of the fourth order, EP4
%plot({0.1*sqrt(0+0.1*s^2),-0.1*sqrt(0+0.1*s^2),0.2*sqrt(0+0.2*s^2),
%-0.2*sqrt(0+0.2*s^2)},s=-12..12,color=black,axes=none,tickmarks=[0,0]);
%
%plot({0.1*sqrt(0+0.1*s^2),-0.1*sqrt(0+0.1*s^2),0.2*sqrt(0+0.2*s^2),
% -0.2*sqrt(0+0.2*s^2)},s=-12..12,color=black,axes=framed,tickmarks=[2,2]);
\begin{figure}[h]                     %instead of \begin{figure}[t]
\begin{center}                         %instead of \begin{center}
\epsfig{file=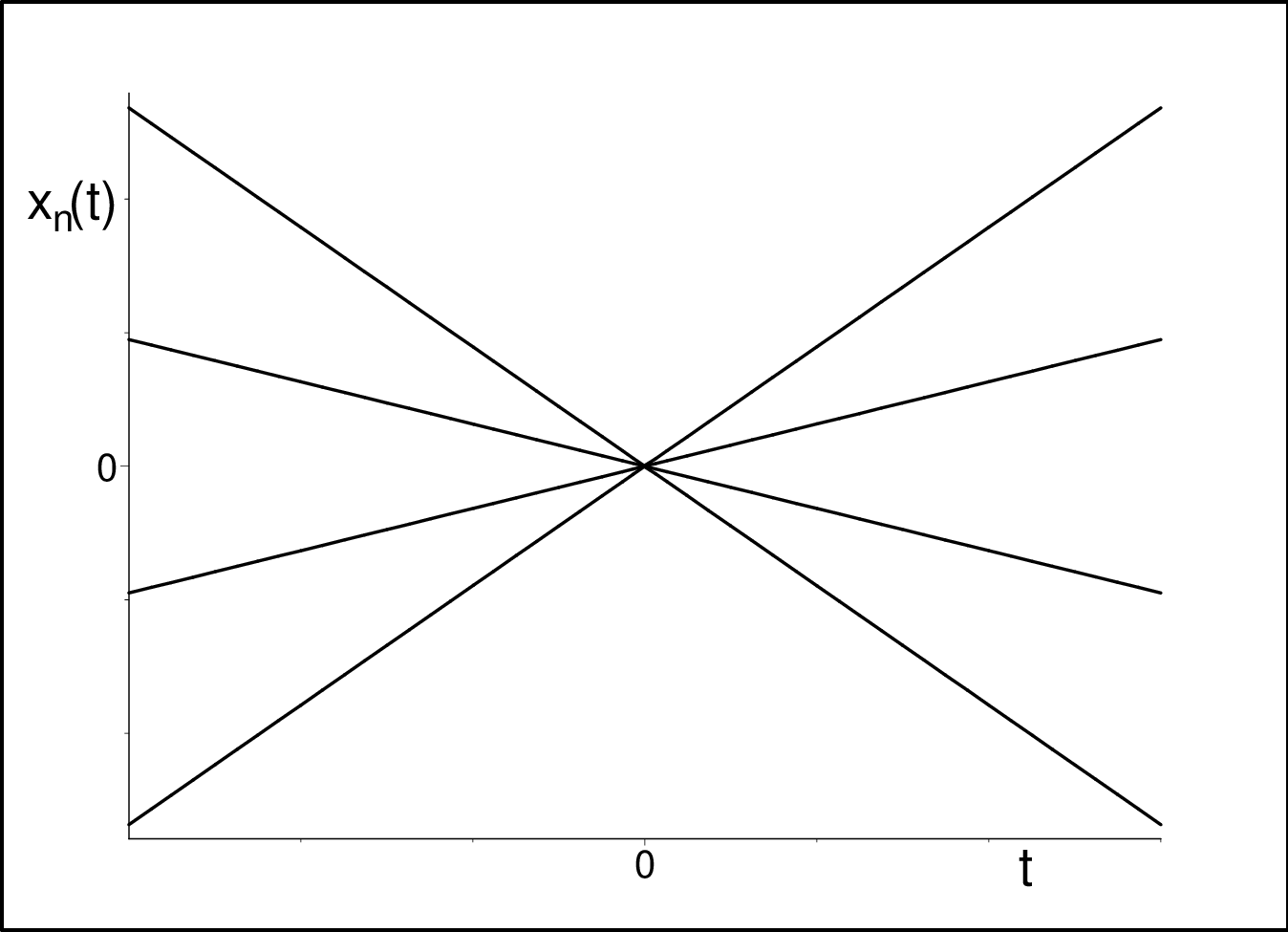,angle=270,width=0.6\textwidth}
\end{center}                         %instead of \end{center}
\vspace{-2mm}\caption{Time-dependence
(i.e., an ``unavoided crossing'') of the
spatial grid points $x_n(t)$ during
the collapse \textcolor{black}{to a single point}
(Big Crunch) and subsequent
re-birth (Big Bang) of the Universe
%in the cyclic cosmology framework.
%
(classical-physics picture, arbitrary units).
 \label{2pic}}
\end{figure}

A comment on the Penrose's eternal cyclic
cosmology hypothesis
has been added
in our recent paper \cite{profri}.
We pointed out there that
before Big Bang,
the two alternative
scenarios as sampled by our present Figures \ref{3pic}
and  \ref{2pic}
are just extremes.
Besides these two
options (i.e., besides the assumption of
no or of an
unchanged observable space at $t<0$, respectively),
it would be possible to consider a certain ``evolutionary''
cosmology in which the
structure of the older \textcolor{black}{Aeon}
can be, vaguely speaking,
``underdeveloped''.

%plot({0.092*sqrt(0+0.23*(-s)^1.99),-0.092*sqrt(0+0.23*(-s)^1.99),
%0.1*sqrt(0+0.1*s^1.99),-0.1*sqrt(0+0.1*s^1.99),0.2*sqrt(0+0.2*s^1.99),
%>  -0.2*sqrt(0+0.2*s^1.99)},s=-12..12,color=black,axes=framed,tickmarks=[2,2]);
\begin{figure}[h]                     %instead of \begin{figure}[t]
\begin{center}                         %instead of \begin{center}
\epsfig{file=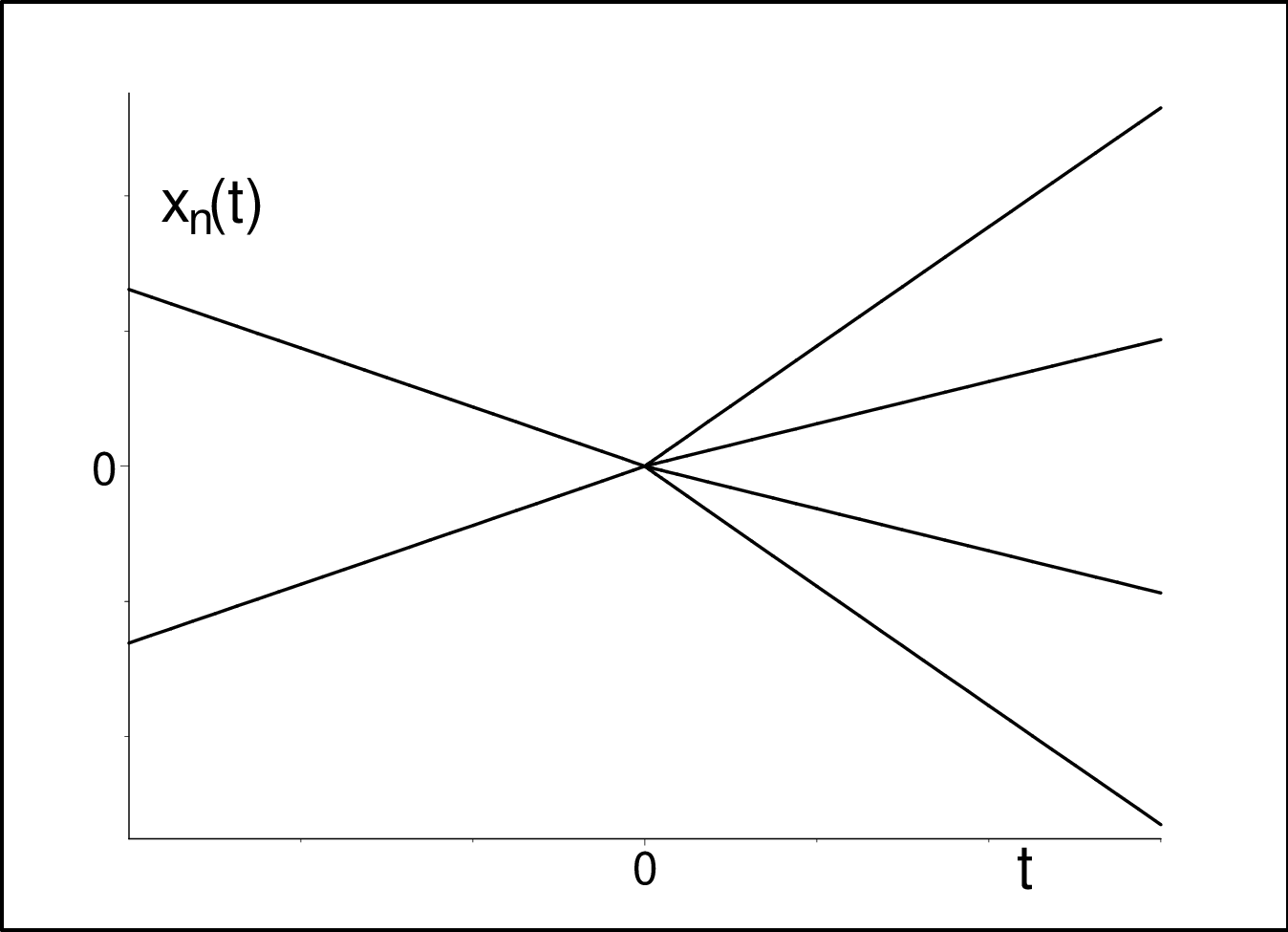,angle=270,width=0.6\textwidth}
\end{center}                         %instead of \end{center}
\vspace{-2mm}\caption{A sample of ``evolutionary'', time-asymmetric
Big Crunch -- Big Bang transition
(classical-physics picture, arbitrary units).
 \label{2bpic}}
\end{figure}

In the language of our schematic pictures,
Figure \ref{2bpic}
can be recalled as an illustration
in which one registers just two ``observable'' grid points at $t<0$.
A more explicit illustration of such an evolutionary
scenario can be found in our older paper
\cite{asymmetry} where we considered
ten observable points at $t>0$.
In our subsequent paper
\cite{asymmetryb}
the schematic ``four-point'' old \textcolor{black}{Aeon}
was studied as
evolving into an``eight-point''
(i.e., strictly twice as large) new \textcolor{black}{Aeon}.

In all of the latter papers
it had to be emphasized that
\textcolor{black}{the transition from classical Big Bang scenarios to
their
quantum mechanical descendants involves
a number of significant conceptual leaps.
The reason is that}
at present,
we do not have any fully consistent quantum
version of the classical Einstein's theory
at our disposal.
Still, even on the present level of our
understanding of the quantized theory,
a prevailing opinion is that
after quantization,
the Big Bang singularity of classical theory
must necessarily be replaced by a regularized mechanism
of a process called ``Big Bounce''
(cf., e.g., \cite{Bounce}
or section 8 in \cite{Rovelli}).
In Rovelli's words, the quantization-related
``absence of singularities'' is in fact
``what one would expect from a quantum theory
of gravity''(see p. 297 in {\it loc. cit.}).

Temporarily dominant as it was, this opinion
has
recently
been challenged by Wang and Stankiewicz
\cite{BBzpet}
who claimed
that in their theory
``the quantized Big Bang is not replaced by
a Big Bounce''.
This was precisely the reason why we
formulated our present project and, in particular, why we
decided to study the problem of
the possible survival of
singularities after quantization
via elementary models.

The presentation of our results will
be preceded by section \ref{dvjesec}
in which we will introduce the basic concepts
and, first of all, the interpretation of Big Bang
as a phenomenon characterized by
the time-dependence of the geometry
of the conventional 3D space.
For our present methodical purposes
the time-dependence of
this geometry
will be simulated,
in a manifestly non-covariant manner,
by the time-dependence of
a few representative spatial grid points $x_n(t)$.

In section \ref{vjesec}
we will accept the hypothesis that
the ``initial'' Big Bang singularity
will not be ``smeared out''
by the quantization
and that its survival can find a consistent
mathematical ground in the Kato's \cite{Kato}
concept of exceptional point (EP).

A more explicit search for consequences
of these assumptions will be started in section  \ref{riplsec}
in which the underlying Hilbert space
of quantum states will be assumed two-dimensional,
and in which
the quantum grid-point operator $X(t)$ representing
(or rather sampling)
the geometry will be just a two-by-two
complex and symmetric matrix.
A key message of our present paper
(viz., a recommendation
that one should pay attention to the singular values
rather than to the eigenvalues of $X(t)$)
finds there its first and most elementary and transparent
illustration.

A series of a few less elementary $N$ by $N$ toy models
will be then analysed in sections \ref{vbtriplsec}
(in which
our illustrative matrices $X(t)$ remain to be complex and symmetric)
and  \ref{btriplsec} (where a different class of toy models
is studied
in order to demonstrate a certain model-independence of our observations
and conclusions).

Our observations will be
thoroughly discussed in section \ref{discussion}
where
we will briefly review some of the basic
challenges encountered during the
attempted quantization of Big Bang
on a more general and example-independent level.
Finally, our results will be briefly summarized in section
      \ref{summary}.

\section{3D space and its few-grid-point representations
\label{dvjesec}}

Our present study was motivated
by the toy models as sampled
by Figure \ref{2bpic}
in which
one expects that after quantization, the time-dependent
positions $x_n(t)$
of the
grid points
should be treated as eigenvalues of
a suitable ``grid point'' operator $X(t)$.
In Figure \ref{2bpic} itself,
only two eigenvalues
remain real
before Big Bang.
The other two eigenvalues are not displayed
because they are not observable.
The corresponding quantum states
(spanning an unphysical part of the Hilbert space)
have to be treated and decoupled as
``ghosts'' \cite{asymmetry}.

In the current physics-oriented literature,
the treatment
of the problem of
ghosts
(i.e.,
of the candidates $X(t)$ for the operators of observables
with the mixed -- real or
real plus complex -- spectra)
is still inconclusive
(cf. \cite{asymmetryb,elitrick}).
Paradoxically, therefore,
our present project of study of the
schematic models
of the
Big Bang singularity
in quantum cosmology
will be mainly guided by the recent
progress achieved by mathematicians.

In this direction, in particular,
Pushnitski with \v{S}tampach \cite{PS1}
opened a new direction of research
by having
proposed a turn of attention
from the information
about the system mediated by the
mixture of the real and complex eigenvalues
(this information is complete but complicated)
to its reduced form represented by the
real quantities called singular values.
% \cite{SV}.

We will follow the guidance.

\subsection{Grid points in conventional Hermitian quantum theory\label{precese}}

%...
Without any reference to the
Penrose's
cyclic-cosmology
\textcolor{black}{realization of the
crossover between Aeons (mediated by a conformal
scaling
and, hence, not quite characterized
by our present Figure \ref{2pic}),}
the
related
transition is,
formally, the simplest possible scenario
\textcolor{black}{in the classical-physics framework}.
The Collapse and Expansion \textcolor{black}{transmutation}
of the 3D space
can be then sampled by
a multiplet of
some arbitrary
representative
time-dependent grid points
(i.e., after the present simplification,
by a
1D
quadruplet $x_n(t) \in {\mathbb{R}}$).

On the classical-physics level,
even
Penrose himself had and has multiple opponents.
Some of them claimed that
we cannot consistently speak about the time before Big Bang
so that, in our present graphical language,
Figures~\ref{2pic} and~\ref{2bpic}
have to be rejected as over-speculative.
Another and, by far, the more relevant counter-argument by the opponents
reflected the
fact
that
due to
the
high-density and
high-temperature
nature of the initial stage of evolution of
our present Universe,
the description of Big Bang must
necessarily be quantum-theoretical.

%avoided crossing:
%near a critical parameter $t = 0$ (schematic
%illustration, arbitrary units).
%%
%plot({0.1*sqrt(1+0.1*s^2),-0.1*sqrt(1+0.1*s^2),0.2*sqrt(1+0.2*s^2),
%-0.2*sqrt(1+0.2*s^2)},s=-12..12,color=black,axes=none,tickmarks=[0,0]);
%
%plot({0.1*sqrt(1+0.1*s^2),-0.1*sqrt(1+0.1*s^2),0.2*sqrt(1+0.2*s^2),
% -0.2*sqrt(1+0.2*s^2)},s=-12..12,color=black,axes=framed,tickmarks=[2,2]);
\begin{figure}[h]                     %instead of \begin{figure}[t]
\begin{center}                         %instead of \begin{center}
\epsfig{file=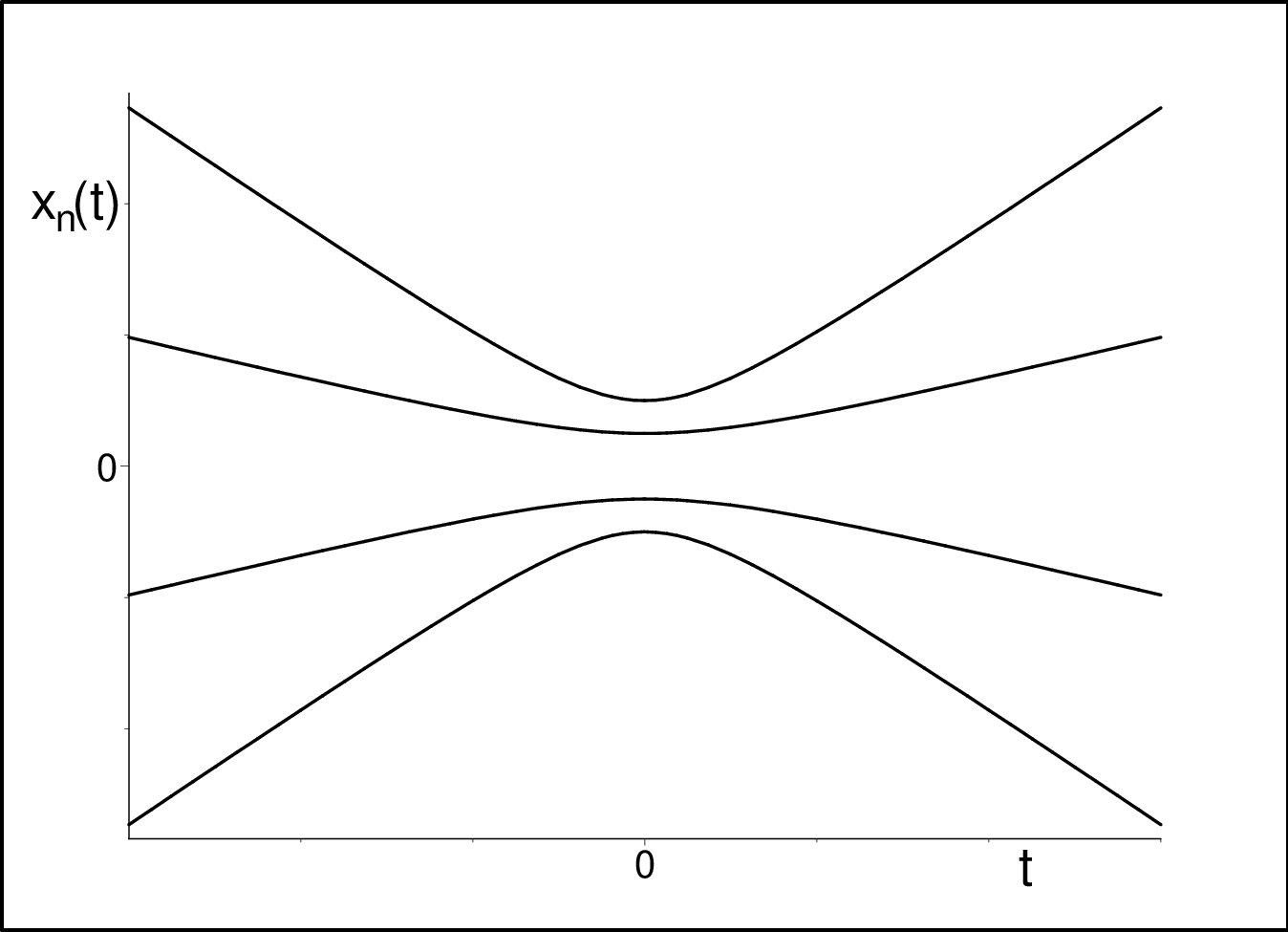,angle=270,width=0.6\textwidth}
\end{center}                         %instead of \end{center}
\vspace{-2mm}\caption{Sample of quantum Big Bounce:
Avoided crossing of
eigenvalues $x_n(t)$ of a self-adjoint grid-point
operator~$X(t)$.
 \label{1pic}}
\end{figure}

%
%\subsection{Grid points in pseudo-Hermitian quantum theory\label{uprecese}}
%

The recent progress
in the quantization of gravity
(cf., first of al,
its loop quantum gravity form \cite{Rovelli,Ashtekar,Thiemann})
could be recalled as a strong
(though not ultimate \cite{BBzpet})
support
of the above-mentioned intuitive
expectation that
the point-like classical
Big-Bang singularity
has to be smeared out
yielding, after quantization, something like Big Bounce~\cite{Bounce}.
In graphical language,
such a disappearance of singularity
can be illustrated by Figure~\ref{1pic}
which samples how the
Big-Bang-related crossing of eigenvalues
happens to become ``avoided''.

\subsection{Grid points in pseudo-Hermitian quantum theory\label{uprecese}}

The
innocent-looking assumption of Hermiticity
$X(t)=X^\dagger(t)$
as accepted in Figure~\ref{1pic}
is of a predictive theoretical relevance.
The point is that the removal
of this assumption
would
reopen
the possibility of existence of
a mathematically fully consistent
point-like form of the Big Bang singularity.

In the abstract framework of quantum theory
one of the most straightforward realizations of
a return to the singularity-admitting evolutions
 %as
 %characterized by
 %Figures \ref{2pic} or \ref{3pic}
is offered by the
so called pseudo-Hermitian
theory \cite{ali,book,Moudro}.
In such a theory
the operators representing observables
can be non-self-adjoint.
One can
build a simplified,
non-covariant model of evolution of the Universe in
which the
space-sampling
grid-point operator is
non-Hermitian,
 $$
 X(t)\neq X^\dagger(t).
 $$
This is the key assumption
which enables one to achieve the simultaneous degeneracy of
all (or at least of some) of
the eigenvalues  of $X(t)$ at the Big Bang time
$t=0$
in a way compatible with the
classical point-like Big Bang hypothesis,
 \be
 \lim_{t\to 0}x_n(t)=x(0)\,
 ,\ \ \ \ n=1,2,\ldots,N\,.
 \label{miku}
 %\forall n\,.
 \ee
In mathematics,
Kato \cite{Kato}
proposed to call such a parameter (i.e., $t=0$
in our time scale) ``exceptional point'' (EP).
In physics, Heiss \cite{Heiss,Heissb} calls its use
``ubiquitous''. In \cite{JCS},
in particular, we explained that
in the quantum
theory of gravity
the tentative point-like EP-interpretation of Big Bang
might prove tenable.

\section{Quantum Big Bang as exceptional point\label{vjesec}}

A number of the related
conceptual as well as technical open problems is enormous \cite{Rovelli}.
In our present paper we felt inspired, in particular,
by the fact that after one
postulates the existence of {\em the same\,} toy-model
grid-point operator $X(t)$ {before and after} the Big Bang,
one immediately has to distinguish between the
two alternative before-the-Big-Bang evolution scenarios
as characterized by our two schematic
Figures \ref{3pic} and \ref{2pic}.

From a more abstract mathematical perspective, the different
choices of
$X(t)$
need not necessarily stay restricted just to one of the latter two
options.
Indeed,
besides
Figure
\ref{3pic}
(in which the $t<0$ spectrum
is all unobservable, complex) and
besides
Figure
\ref{2pic}
(in which the whole $t<0$ spectrum
is assumed real),
one can easily imagine the existence of
many other $t<0$ spectra
which may only be partially real (cf. Figure \ref{2bpic}).

We have hardly any chances
of testing
the hypotheses
concerning the properties
of the Universe during the pre-Big-Bang ``\textcolor{black}{Aeon}''
(but no certainty -- see \cite{Penroseb}).
One could even feel skeptical
concerning the reliability of any
experiment-based insight into the structure
of the 3D space shortly after the  {\em current\,} Big Bang.
Thus, one cannot exclude
a survival of some complex
(i.e., unobservable)
eigenvalues of $X(t)$ at any time $t<0$
or $t>0$.

The  requirement of an extension
of the class of
the hypothetical cosmological quantum
models beyond the two extreme simplifications
sampled by
Figures
\ref{3pic} and
\ref{2pic}
is also well motivated
by a broader
mathematical experience and
physical context: {\it Pars pro toto},
let us recall paper  \cite{Uwe},
a typical many-body model
in which the authors
worked
with non-Hermitian operators.

\subsection{Alternative unfolding patterns \label{advjesec}}

Before we
proceed to the formulation of our project
let us note that
what remains unclear is the very
applicability of a consistent quantization
of
such a
generalized approach to the Big Bang problem.

In this setting
and, in particular, in the cosmological Big Bang context,
the use of exactly solvable models can be found indispensable.
Although the detailed analysis of such models
need not necessarily lead to
phenomenological and experimentally verifiable
predictions,
it can certainly complement at least some of
the existing cosmological speculations
about the issues like
an observability of the traces of
what happened before Big Bang \cite{Penrose,Penroseb},
or like
the question of existence or absence of a
Big Bang singularity in cosmology
after quantization \cite{Bounce,BBzpet,piech}.

During the purely mathematical
study of dynamics of a toy-model quantum
Universe near its hypothetical EP-related
singularity (located at $t=0$ in our pictures)
we might wish {\em not\,} to distinguish
between the $t>0$ evolution
{\em after\,} the Big Bang and the $t<0$ evolution
{\em before\,} the Big Bang.
Nevertheless,
a necessary return from mathematics
to physics
would immediately force us to
see a deep contrast
between our ability of
making experiments in
the current or in the preceding \textcolor{black}{Aeon}.

Although the strength of this contrast has
partially been denied by Penrose et al \cite{Penrose},
the natural asymmetry between the pre-Bang and post-Bang
era (i.e., in the Penrose's terminology, \textcolor{black}{Aeon}),
it might make sense to keep it implicitly in mind even
during our forthcoming
methodical coverage of both of these regimes
on a more or less equal footing.

Even under
such \textcolor{black}{an} idealization
of possible physics behind
our present toy-models of the Big Bang
(admitting the loss of the reality of the spectrum
at both signs of the time $t$),
we are not going to classify and study the general cases,
having in mind that our present methodical purposes
will be well served even when we
choose, for our explicit illustrative
calculations, just the models in which
the spectrum would have  most general
structure of Figure \ref{2bpic}.

A decisive methodical advantage of the latter
choice is that the models will
support {\em both\,} the real and complex eigenvalues.
Still,
our present observations will be
far from conclusive.
Long before their possible
completion in the future,
our methodically motivated
simulations of the quantum Big Bang event
may help even though they were only made feasible after their
truly drastic simplifications.

\subsection{Vicinity of quantum Big Bang}

In our recent study \cite{axioms}
we
pointed out that
one of the eligible consistent
descriptions of a
point-like
quantum
Big Bang
(i.e., of a strictly quantum analogue of Figure~\ref{3pic} or \ref{2pic})
could be
provided via the
pseudo-Hermitian reformulation of
conventional
quantum theory
(see, e.g., review \cite{ali}).
Such a
description has to be based on an identification
of the Big Bang instant $t=0$
with the Kato's exceptional-point parameter (EP, \cite{Kato}).
In the manner supported by an illustrative example
we demonstrated
that
such an identification
could be feasible.
Now we plan to support the claim by
its deeper technical analysis.

% Big Bang and observability puzzle

From a broader conceptual point of view the main weakness of the EP-based
picture of the birth of the Universe is that the support of its validity
remained restricted to the specific subclass of
exceptional points, namely, to those evolution patterns which can be
characterized, in our intuitive graphical presentation,
by Figures \ref{3pic} or \ref{2pic}.
This strictly requires that
{\em after\,} the Big Bang event (i.e., at $t>0$),
{\em all\,} of the eigenvalues
$x_n(t)$ of an admissible physical
grid-point
operator~$X(t)$
remain real.
Simultaneously, {\em all or none\,} of them
have to be real at $t<0$, i.e.,
{\em before\,}
the Big Bang event.

Even in the context of pure mathematics, both of the
respective
latter postulates
(viz., of the Penrose's ``cyclic re-birth'', or
of a more pragmatic hypothesis of ``nothing before Big Bang'')
must be considered
over-restrictive.
A reduction of the picture to these two extremes
can hardly be considered satisfactory also
from a purely phenomenological perspective.
Indeed, we may recall, for comparison,
the situation in many-body physics.
In the framework of Bose-Hubbard model,
the study of exceptional points
by
Graefe et al \cite{Uwe}
revealed
that although an extreme
scenario as sampled here by Figure \ref{3pic}
does exist, its occurrence only represents a very small
fraction of all of the mathematically admissible
alternatives.
The latter authors observed that in
spite of the fairly realistic background of their model,
the emergence of a
complex part in any eigenvalue
can hardly be excluded.
In
the vicinity of
a generic EP, in general.
some of the eigenvalues happened to remain real
(i.e., observable) and some of them not.

The same mixed spectra can be expected to occur also in
the EP-based cosmological applications.
The more so
near the quantum Big Bang
where the grid-point operators $X(t)$ are
introduced, in most cases, via a purely
pragmatic and physics-oriented process.
Hence, the complexity of at least some of the eigenvalues
should be perceived as generic.

In the vicinity of the EP parameter $t=0$,
as a consequence,
the quantum system in question
(i.e., in our case, the quantum Universe)
will have to be
treated
by the methods which are usually
used for the description of the open systems
with resonances \cite{Nimrod}.
Having this in mind
we will study the general EP-unfolding process via several
elementary but non-numerically tractable
toy models.

\subsection{Unobservable grid points}

In paper \cite{Uwe}
the model has been kept
simple while still
admitting the non-real elements in the
spectrum.
In the most conventional context of quantum mechanics
of many-body systems
the authors worked with
a parameter-dependent non-Hermitian Hamiltonian
yielding both the bound and resonant-states.
They were able to
study and describe the parameter-dependence of the spectra
using standard
mathematical techniques including
not only the numerical but also
the analytic and perturbative ones.

The feasibility of
these
techniques
will decrease with an increase of the complexity
of the operator. Once we plan to
turn attention of the quantum cosmology and Big Bang,
the acceptance of some truly drastic
simplifications
seems necessary.

Even \textcolor{black}{though} we decided to mimic just
the passage of the Universe through its
spatial singularity using a
grid-point toy-model operator $X(t)$,
we will have to
treat the time as a parameter.
What we gain is that the dynamics of any underlying ``spatial background''
becomes quantized.
What we lose are the chances \textcolor{black}{of having}
our picture of reality covariant.

Although we %have to
insist on the reality
of $x(0)$ in Eq.~(\ref{miku}),
\textcolor{black}{we will assume just}
a partial observability of
the Universe near Big Bang.
\textcolor{black}{This will enable us to
get a certain source of the
observational signatures which could distinguish
our present, EP-based
quantum Big Bang scenario from
its various conventional alternatives.
Thus, in a small
but, in principle, both-sided vicinity of $t=0$}
we will
admit the complexity
of at least
some of the eigenvalues of $X(t)$,
  $$
   \exists \,n\,,t \ \ \ {\rm that}\ \ \ x_n(t) \in {\mathbb C}\,.
   $$
For virtually any
sufficiently realistic and general choice of $X(t)$,
the latter assumption would
make the evaluation of its spectrum
complicated.
This is a very core of the problems which will be addressed
in what follows.

\section{Two-grid-point simulation of Big Bang\label{riplsec}}

A part of the reason why we intend to reduce
the study of the eigenvalues $x_n \in \mathbb{C}$
to the study of singular values $\sigma_n \in \mathbb{R}$
can be seen in the rather serious technical obstacles
encountered during the numerical localization
of complex eigenvalues. The most immediate insight
in some properties of this reduction
can be provided when we replace
the operators by matrices.
For a deeper understanding of
correspondence between the physics of
evolution of the Universe near Big Bang and
the mathematics of unfolding of the
EP-based degeneracies of the quantum grid-point spectra
it makes sense to
mimic these processes using a few most elementary toy models.

\subsection{Grid-point operator $X(t)$: Complex symmetric choice}

% SV - BH
%
%plot({0.1*sqrt(0+0.1*s^1.99),-0.1*sqrt(0+0.1*s^1.99),0.2*sqrt(0+0.2*s^1.99),
% -0.2*sqrt(0+0.2*s^1.99)},s=-6..12,color=black,axes=framed,tickmarks=[2,2]);
%
%%
%
\begin{figure}[h]                     %instead of \begin{figure}[t]
\begin{center}                         %instead of \begin{center}
\epsfig{file=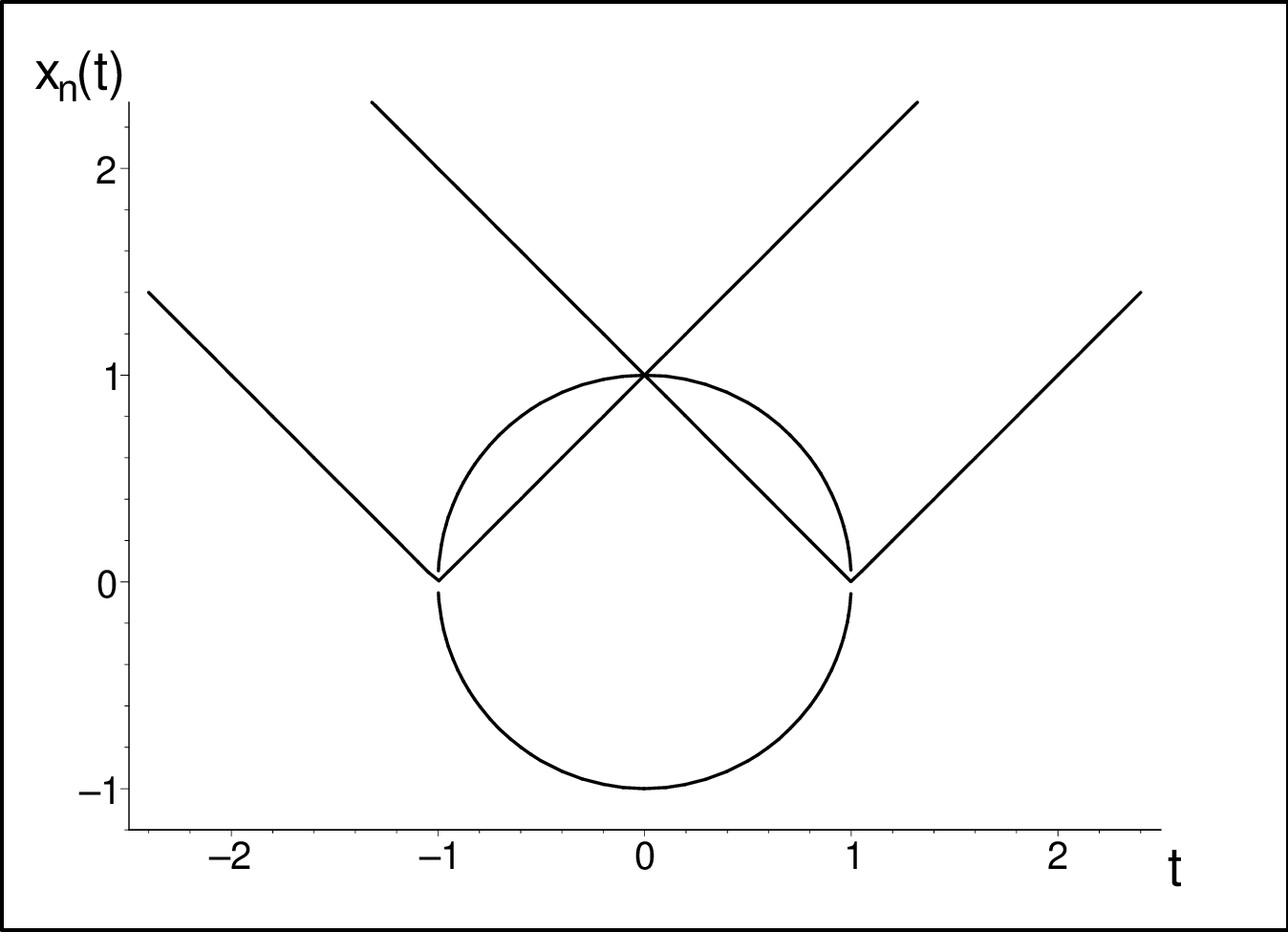,angle=270,width=0.6\textwidth}
\end{center}                         %instead of \end{center}
\vspace{-2mm}\caption{A comparison of the time-dependence of the
real
eigenvalues $x_n(t)$ (= the circle)
with the time-dependence of the singular values $\sigma_n(s) $
(= straight lines)
for the two-by-two non-Hermitian toy-model
matrix
(\ref{ufo}).
One has a choice here between the
Big Bang scenario at
$t^{(EP)}_{(BB)}=-1$
and the
Big \textcolor{black}{Collapse} scenario
at $t^{(EP)}_{(BC)}=+1$.
 \label{5pic}}
\end{figure}

 \noindent
A particularly transparent explicit illustration
of the mechanism of the EP-based unavoided crossings
of the time-dependent eigenvalues $x_n(t)$
of the time-dependent and manifestly non-Hermitian
operators $X(t)$
is offered by the following two-by-two-matrix
grid-point-operator toy model
 \be
 {X}(t)=\left[ \begin {array}{cc} -it&1\\
 \noalign{\medskip}1&it\end {array} \right]
 \label{ufo}
 \ee
which is easily
diagonalized,
 $$
 {X}(t) \to
 \mathfrak{\xi}(t)=\left[ \begin {array}{cc} \sqrt {1-{t}^{2}}&0\\
  \noalign{\medskip}0&-\sqrt {1-{t}^{2}}\end {array} \right]\,.
 $$
The two-level
spectrum
of such a model
is strictly real if and only if $t \in [-1,1]$ (see the circle in
Figure \ref{5pic}).
This is a closed interval
but its endpoints are
the Kato's
exceptional points.
They are manifestly unphysical because
matrix (\ref{ufo})
ceases to be diagonalizable
in the limit of $t \to \pm 1$.

Such an observation enables us to
treat the matrix of Eq.~(\ref{ufo})
as a highly schematic model of
the two-grid-point Universe
in two ways.
In its first, ``Big Bang'' interpretation
(which would be conceptually analogous to Figure \ref{3pic} above)
we may localize the Big Bang event at $t^{EP)}_{(BB)}=-1$,
and we will have to keep our ``age of the Universe''
well below the origin
in suitable units, $-1 < t \ll 0$.

In another, complementary ``Big  \textcolor{black}{Collapse}'' interpretation
of the model,
we could localize another, ``inverse''
Big  \textcolor{black}{Collapse} event at $t^{EP)}_{(BC)}=+1$
while keeping the  time of the ``end of life of the preceding \textcolor{black}{Aeon}''
well separated from the origin, with, perhaps, $0 \ll  t < 1$
in some
other suitable units.

% SV - BH
%
%plot({0.1*sqrt(0+0.1*s^1.99),-0.1*sqrt(0+0.1*s^1.99),0.2*sqrt(0+0.2*s^1.99),
% -0.2*sqrt(0+0.2*s^1.99)},s=-6..12,color=black,axes=framed,tickmarks=[2,2]);
%
%%
%
\begin{figure}[t]                     %instead of \begin{figure}[t]
\begin{center}                         %instead of \begin{center}
\epsfig{file=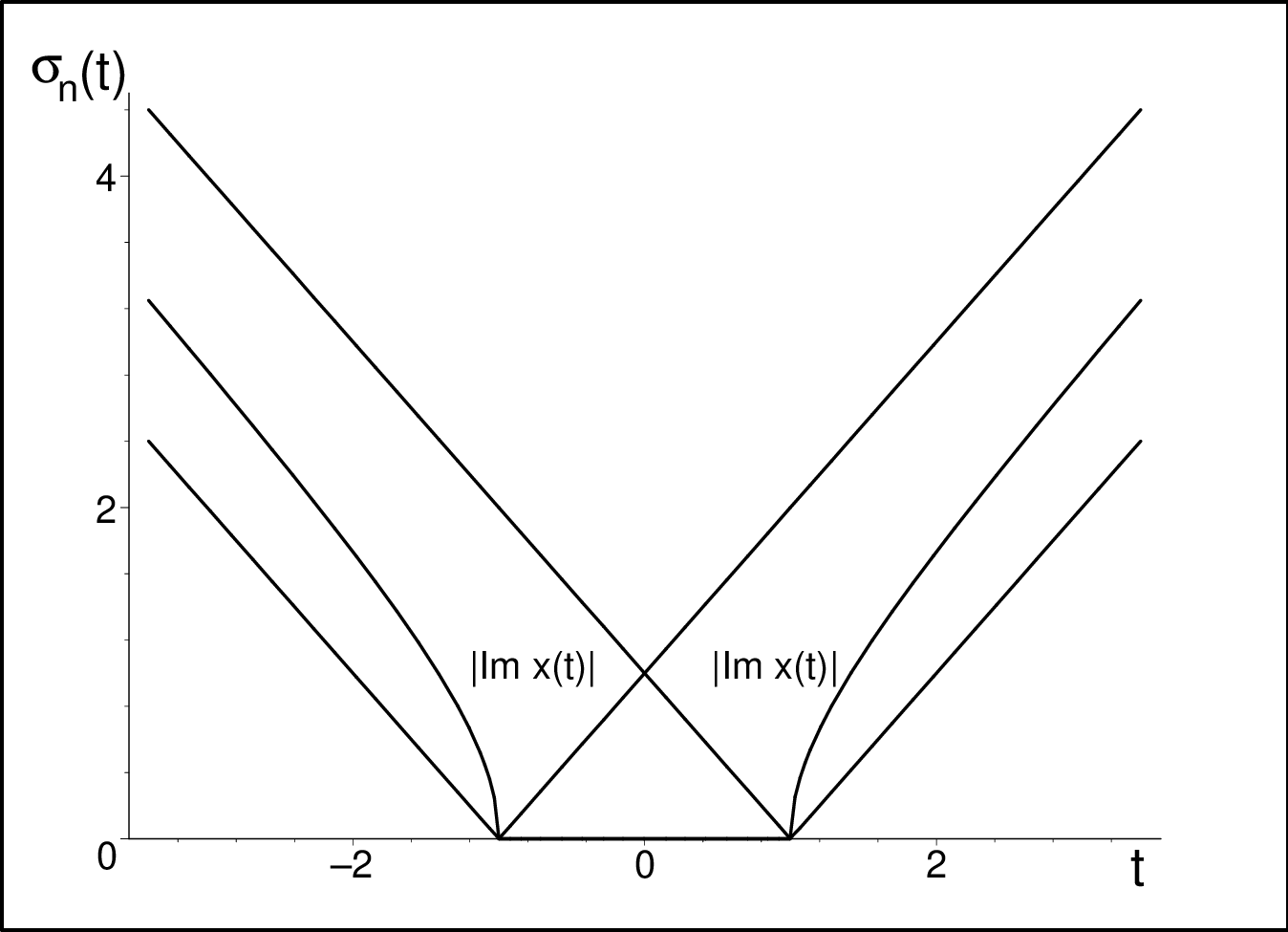,angle=270,width=0.6\textwidth}
\end{center}                         %instead of \end{center}
\vspace{-2mm}\caption{The phenomenon of bracketing
of the absolute values of the purely imaginary eigenvalues $x_n(t)$
of matrix
(\ref{ufo})
by the pairs of singular values $\sigma_n(t) $ of the same
non-Hermitian matrix at $|t|\geq 1$
(i.e., in the unstable-solution dynamical regime).
 \label{u5pic}}
\end{figure}

Along the rest of the real line of time
(and, in both \textcolor{black}{Aeons}, sufficiently far from the origin,
i.e., for $|t|>1$)
the grid-point spectrum of the model becomes purely imaginary:
See the hyperbolic-curve
graph of $|{\rm Im\ }x_n(t)|$
in
Figure~\ref{u5pic}.
This would make  the 3D space, naturally,
hardly observable, in the conventional current perception
of the theory at least.

\subsection{The role of singular values}

In
contrast to the curves
representing the spectra, the singular
values  $\sigma_n(t) $
are, in
our pictures,
real, linear and non-negative functions of time at all
$ t \in (-\infty,\infty)$.
Once we
restrict attention to the upper halves of the spectral curves,
a quick inspection of our two pictures reveals that
every such a curve
lies inside a strip defined by singular values.
In other words, it is
minimized and maximized
(we may say ``bracketed'') by the two
singular-value lines.

In particular,
inside the ``observable \textcolor{black}{Aeon}'' interval of $t \in (-1,1)$,
Figure \ref{5pic} indicates that
the
bracketing
applies to
the real and positive eigenvalue
(plus, after the
mere change of sign, also to the other, negative eigenvalue).
For
$t \notin (-1,1)$ (i.e., far from the origin),
the other
Figure \ref{u5pic} shows how
the same bracketing
applies to the absolute value of the
complex eigenvalue.
The (in principle, simpler)
evaluation of singular values
appears useful,
for the purposes of a rough spectral estimates at least.

Various forms of
the bracketing properties of the
(always real)
singular values is
well known in mathematics.
% \cite{SV}.
In the context of physics one should only keep in mind that
strictly speaking,
our most elementary toy-model
operator $X(t)$ of Eq.~(\ref{ufo})
loses its status of an observable
at
$t \notin (-1,1)$.
Still, there exist phenomenological reasons
due to which even the complex eigenvalues $x_n(t)$
have to be evaluated and reinterpreted, say,
as characteristics of an
unstable
quantum system.

The closest analogy of such a situation
can be found in the emergence of unstable
resonances in the so called open quantum systems, say,
in the nuclear, atomic or molecular physics~\cite{Nimrod}.
A return to stability as
encountered at $t \in (-1,1)$
can be then perceived
as a physics-specifying emergence of
an observable spatial position in our toy model.
A return to
the reality of the
eigenvalue implies a return to
observability of
grid points.
Once we have to predict
an outcome of a measurement, we have to localize
them as precisely as possible.
For this reason,
the bracketing
may play the role of a rough estimate.

The
complexity of the eigenvalues is often interpreted as a
more or less unavoidable consequence of
an
influence of an (in principle, unknown)
``environment''.
Still, even if one accepts such a philosophy as ``more realistic'',
another obstacle emerges from the necessary generalization
of the underlying mathematics.
Whenever the eigenvalues happen to be complex,
their practical numerical evaluation becomes
technically much more complicated.

This is the reason why, in practice, the localization
of complex eigenvalues
is often being replaced by the evaluation
of singular values. The
process is simpler because the singular values
remain real. Last but not least, their definition
is comparatively easy. % (cf. \cite{SV}):
For any non-Hermitian matrix $Q$
they
coincide with the non-negative square roots
of eigenvalues of an associated, manifestly
Hermitian matrix product $Q^\dagger\,Q$.

It is worth adding that our toy model matrix (\ref{ufo})
can be interpreted as
parity-time symmetric {\it alias\,}
${\cal PT}-$symmetric \cite{BG,BB}
{\it alias\,} Krein-space Hermitian \cite{Langer,Kuzhel}.
Due to such an additional symmetry,
the form (i.e., the time-dependence)
of its singular values $\sigma_n(t)$
is simpler than
that of the eigenvalues $x_n(t)$
themselves.

In order to see this, we only have to recall the
definition by which the singular value of $
 {X}(t)$
is just a (plus-sign) square root of an eigenvalue
of the product
 $$
 \mathbb{{X}}(t)={X}^*(t)\,{X}(t)
 =\left[ \begin {array}{cc} {t}^{2}+1&2\,it\\
 \noalign{\medskip}-2\,it&{t}^{2}+1\end {array} \right]\,.
 $$
The diagonal isospectral partner of the latter matrix is easily
evaluated,
  $$
  \widetilde{\mathbb{{X}}}(t)
  =\left[ \begin {array}{cc} {t}^{2}+1+2\,t&0\\
  \noalign{\medskip}0&{t}^{2}+1-2\,t\end {array} \right]\,.
  $$
Although the spectrum of ${X}(t)$ becomes complex
at $|t| > 1$, the singular values
of the same operator remain both real
along the whole real axis of $t$.

For illustration,
the comparison of the two alternative characteristics
of model (\ref{ufo})
is displayed in  Figure \ref{u5pic}.
One discovers there a non-trivial
relationship between the eigenvalues and
singular values in the unstable-state dynamical regime.
The picture demonstrates that the knowledge of
the singular values $\sigma_n(s)$
offers in fact a truly non-trivial information
about the spectrum of the matrix
even when this spectrum is complex.

Another, serendipitous benefit of the model is that
the elementary
time-dependence
of our grid-point matrix $X(t)$ of Eq.~(\ref{ufo})
appears reflected by the equally elementary (viz., linear)
time-dependence
of
its singular values $\sigma_n(t)$.
In our
Figures \ref{5pic} and
\ref{u5pic} the singular values
appear as straight lines, therefore.

A climax of the story can be seen in the
fact that
in contrast to the
user-friendly scenario with the real spectrum at $t \in (-1,1)$,
the phenomenon of bracketing
keeps playing a useful role at $t \notin (-1,1)$.
According to Figure \ref{u5pic}, the
bounds imposed upon the imaginary parts of the eigenvalues
as
provided by the singular values
look, for practical purposes, impressive.

Although the spectrum-bracketing property
as sampled by Figure \ref{u5pic}
may look like an artifact emerging in a specific model,
one could recall the existing mathematics behind the
singular values
in order to find a general theory of
the related inequalities.
%
%(see also the
%list of further references in \cite{SV}).

The only remaining question is
whether such an observation of bracketing is not
too much model-dependent. Independent tests
are necessary.

\section{More grid points\label{vbtriplsec}}

%, singular values and bracketing
%
%Tracing the trends: Complex symmetric matrices
%
%Increase of the number of grid points:
%
%Tracing the trends: Complex six by six model
%
%complex vs real
%
%complex easier, real analogous (numerically more challenging, left to the readers)

%\section{Singular values

For the majority of the larger $N$ by $N$ matrices
the search for the
eigenvalues and singular values
is a purely numerical task in general.
Still, there are exceptions.
For illustration let us first recall the $N$ by $N$
matrices
of paper \cite{tridiagonal}.

%\subsection{Complex but symmetric $X(t)$\label{hyjesec}}

Nothing truly new emerges when one moves from the
above-considered matrices of dimension $N=2$
to the models with larger $N$ (for the sake of definiteness
we may and will keep these dimensions even).
We will see that the evaluation of the singular values
and, in particular, their bracketing property can still
offer a fairly nontrivial
insight in the structure
and dynamics of the quantum system in question.

\subsection{Four-grid-point simulation of Big Bang\label{atrip}}

The abstract and general mathematical
statements
taken from the literature
can be complemented
by their very explicit algebraic and graphical
support.
Its first version is provided by the
closed-form solvability of
the following four-by-four
grid-point-operator
matrix
 \be
 {X}(t)= \left[ \begin {array}{cccc}
  -3\,it&\sqrt {3}&0&0\\\noalign{\medskip}
 \sqrt {3}&-it&2&0\\\noalign{\medskip}0&2&it&\sqrt {3}
 \\
 \noalign{\medskip}0&0&\sqrt {3}&3\,it\end {array} \right]
 \label{bufom}
 \ee
and by its unitary-equivalent
diagonalized
avatar $\mathfrak{\xi}_{}(t)$ with four non-vanishing elements
 \be
 \mathfrak{\xi}_{11}(t)=\sqrt {1-t^2}=-\mathfrak{\xi}_{44}(t)\,,\ \ \
 \mathfrak{\xi}_{22}(t)=3\,\sqrt {1-t^2}=-\mathfrak{\xi}_{33}(t)\,.
 \label{brac4}
 \ee
Such a
\textcolor{black}{structure of the model
seems to suggest
a suitable underlying symmetry.
Unfortunately, any such a symmetry (or, perhaps,
another  elegant explanation
of the solvability) is not know yet.
In Eq.~(\ref{brac4}),
the nice off-diagonal elements
emerged as a result of a brute-force
computer-assisted algebra as presented in the 2007 paper  \cite{tridiagonal}.
Anyhow, the}
result
yields, after a suitable rescaling of time
near $t^{(EP)}_{(BB)}=-1$, just
an explicit
realization of
the generic four-level spectrum as sampled
in Figure \ref{3pic} above.
Its Big  \textcolor{black}{Collapse} alternative
near $t^{(EP)}_{(BC)}=+1$
is obvious
so that it need not be discussed separately.

%
%%Big Bang 2 (including inflation):
%%
%%plot({0.1*sqrt(0+0.1*s^1),-0.1*sqrt(0+0.1*s^1),0.2*sqrt(0+0.2*s^1),
%% -0.2*sqrt(0+0.2*s^1)},s=-6..12,color=black,axes=framed,tickmarks=[2,2]);
%%
%%,spectrum = real at $s \leq 0$,
%% exceptional point of the fourth order, EP4
%%
%%%
%\begin{figure}[h]                     %instead of \begin{figure}[t]
%\begin{center}                         %instead of \begin{center}
%\epsfig{file=picol.eps,angle=270,width=0.6\textwidth}
%\end{center}                         %instead of \end{center}
%\vspace{-2mm}\caption{Big-Bang-simulating
%scenario with inflation (arbitrary units).
% \label{4pic}}
%\end{figure}

\begin{figure}[t]                     %instead of \begin{figure}[t]
\begin{center}                         %instead of \begin{center}
\epsfig{file=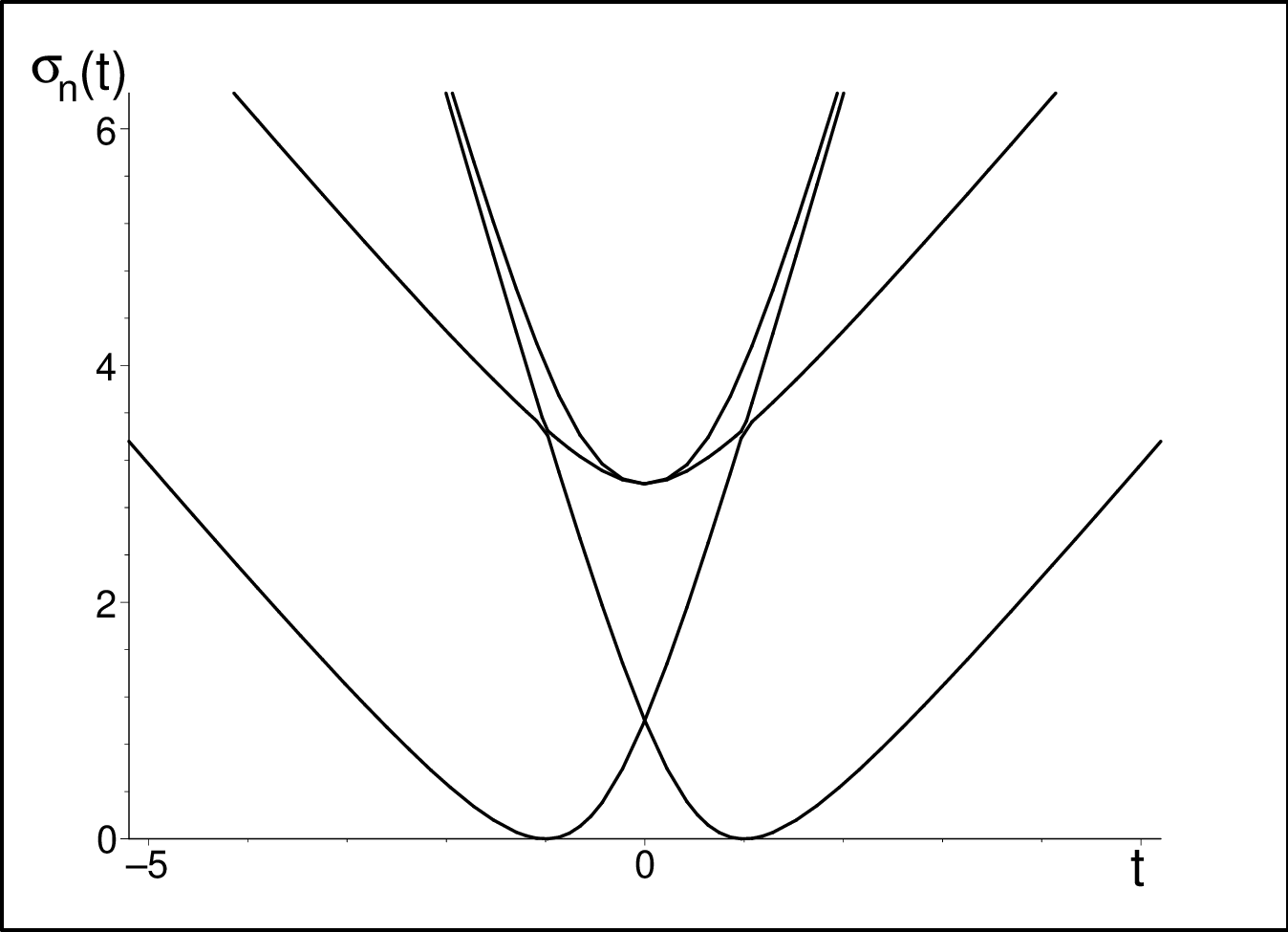,angle=270,width=0.6\textwidth}
\end{center}                         %instead of \end{center}
\vspace{-2mm}\caption{Singular values
for the complex symmetric
four-by-four matrix~(\ref{bufom}).
 \label{55pic}}
\end{figure}

Once we move to the study of singular values, we find it
easy to form the Hermitian-matrix product
 $$
 \mathbb{{X}}(t)={X}^*(t)\,{X}(t)=
  \left[ \begin {array}{cccc} 9\,{{t}}^{2}+3&2\,i{t}\sqrt {3}&2\,\sqrt {3}&0
\\\noalign{\medskip}-2\,i\sqrt {3}{t}&7+{{t}}^{2}&4\,i{t}&2\,\sqrt {3}
\\\noalign{\medskip}2\,\sqrt {3}&-4\,i{t}&7+{{t}}^{2}&2\,i{t}\sqrt {3}
\\\noalign{\medskip}0&2\,\sqrt {3}&-2\,i\sqrt {3}{t}&9\,{{t}}^{2}+3
\end {array} \right]
 $$
and to find the quadruplet
of its eigenvalues,
  $$
  \sigma_{\pm,\pm}(t)=
  5+5\,{{t}}^{2} \pm 2\,{t} \pm 4\,\sqrt {1-{t}-{{t}}^{3}+{{t}}^{4}}\,.
  $$
A better insight in the shape of these four curves in provided by
Figure \ref{55pic}.
In contrast to the preceding two-grid-point case, their time-dependence
ceased to be linear. Still, the
bracketing role of these curves remained analogous
to the one displayed in Figures \ref{5pic} and~\ref{u5pic}
above.

For a verification of the latter observation,
interested readers could recall the elementary formula
(\ref{brac4}) an insert the real spectrum (= rescaled circles)
and/or the absolute values of the imaginary grid points
(= rescaled hyperbolas) into Figure \ref{55pic}.
This insertion may be also accompanied
by the analogous
explicit comments
concerning the numerical quality of the bracketing.

Another, less obvious observation might concern the
closed form of the
results. Although
the exact formulae for the
spectrum itself would survive, in our particular model,
at any finite number $N$ of grid points (cf., e.g., \cite{grid}),
it is not known whether
the time-dependence of the related singular values
remains equally explicit or, at least,
qualitatively predictable.
This is to be tested now
by an explicit computation in what follows.

\subsection{Six by six matrix $X(t)$\label{6ubtriplsec}}

Grid-point operator
 \be
 {X}(t)= \left[ \begin {array}{cccccc} -5\,i{t}&\sqrt {5}&0&0&0&0\\
 \noalign{\medskip}\sqrt {5}&-3\,i{t}&2\,\sqrt {2}&0&0&0
\\\noalign{\medskip}0&2\,\sqrt {2}&-i{t}&3&0&0\\\noalign{\medskip}0&0&3&
i{t}&2\,\sqrt {2}&0\\\noalign{\medskip}0&0&0&2\,\sqrt {2}&3\,i{t}&\sqrt {5
}\\\noalign{\medskip}0&0&0&0&\sqrt {5}&5\,i{t}\end {array} \right]
 \label{bufor}
 \ee
is still easily
diagonalized yielding the spectrum,
$$
\mathfrak{\xi}_{11}(t)=5\,\sqrt {1-{t}^2}=-\mathfrak{\xi}_{66}(t)\,,\ \ \
\mathfrak{\xi}_{22}(t)=3\,\sqrt {1-{t}^2}=-\mathfrak{\xi}_{55}(t)\,,\ \ \
\mathfrak{\xi}_{33}(t)=\sqrt {1-{t}^2}=-\mathfrak{\xi}_{44}(t)\,.
$$
A fully analogous formula for spectrum holds, after all, at any number of grid points $N$
(see \cite{passage}).

Now, the first real challenge comes with the search for the singular values
at $N=6$. Once they are defined via eigenvalues of pentadiagonal matrix
 $$
 \mathbb{{X}}(t)={X}^*(t)\,{X}(t)=
  \left[ \begin {array}{cccccc} 25\,{{t}}^{2}
  +5&2\,i{t}\sqrt {5}&2\,\sqrt {5}\sqrt {2}&0&0&0\\
  \noalign{\medskip}-2\,i{t}\sqrt {5}&13+9\,{{t}}^{2}&4\,
i{t}\sqrt {2}&6\,\sqrt {2}&0&0\\\noalign{\medskip}2\,\sqrt {5}\sqrt {2}&
-4\,i{t}\sqrt {2}&17+{{t}}^{2}&6\,i{t}&6\,\sqrt {2}&0\\\noalign{\medskip}0&6
\,\sqrt {2}&-6\,i{t}&17+{{t}}^{2}&4\,i{t}\sqrt {2}&2\,\sqrt {5}\sqrt {2}
\\\noalign{\medskip}0&0&6\,\sqrt {2}&-4\,i{t}\sqrt {2}&13+9\,{{t}}^{2}&2\,
i{t}\sqrt {5}\\\noalign{\medskip}0&0&0&2\,\sqrt {5}\sqrt {2}&-2\,i{t}
\sqrt {5}&25\,{{t}}^{2}+5\end {array} \right]
 $$
the goal is only achieved using a computer and a symbolic manipulation
software yielding the six real and non-negative eigenvalues in the form of
the well known Cardano formula.

\begin{figure}[t]                     %instead of \begin{figure}[t]
\begin{center}                         %instead of \begin{center}
\epsfig{file=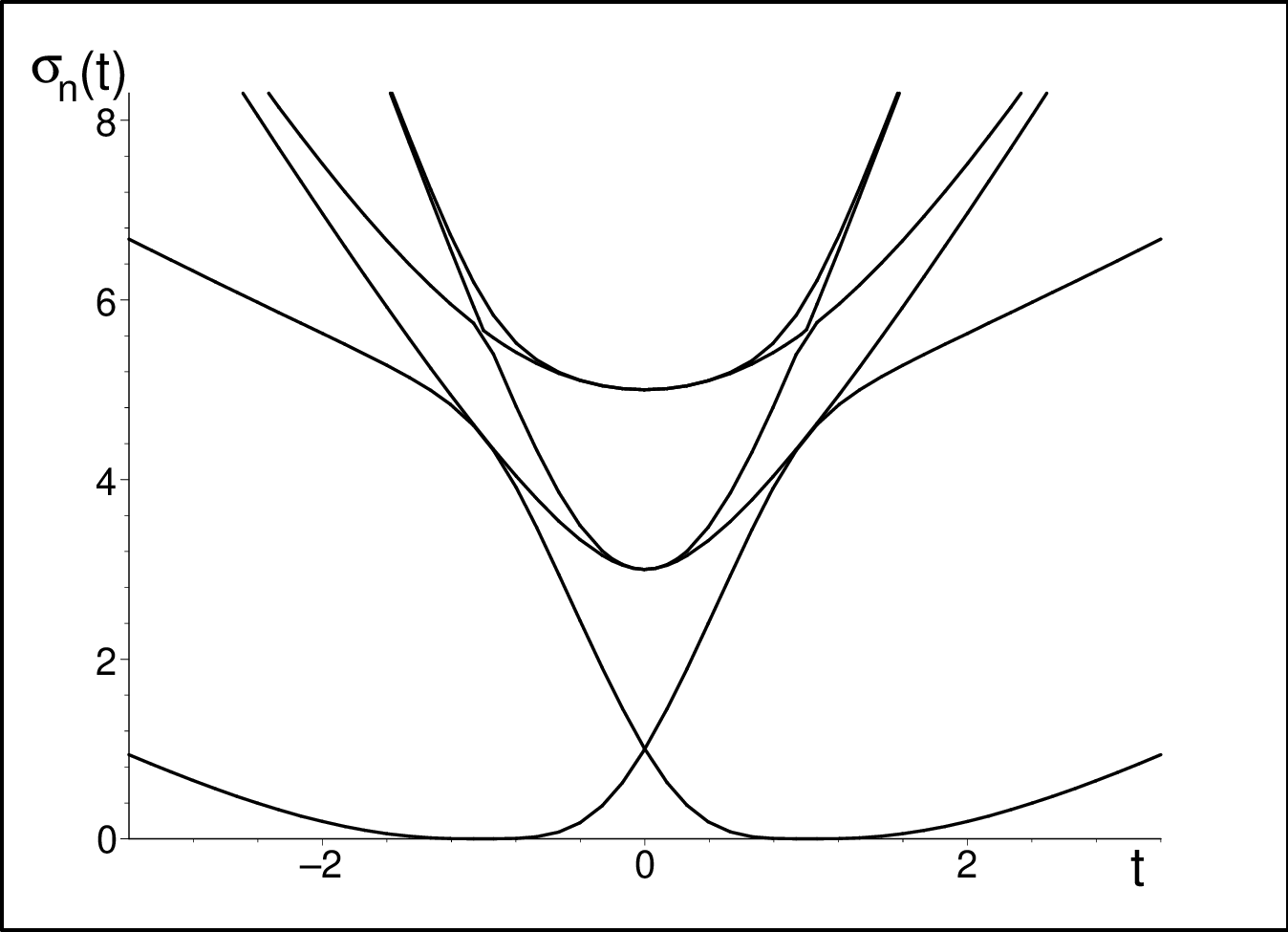,angle=270,width=0.6\textwidth}
\end{center}                         %instead of \end{center}
\vspace{-2mm}\caption{Singular values
for the complex symmetric
six by six matrix~(\ref{bufor}).
 \label{s55pic}}
\end{figure}

This means that the result is still available in closed form, but
the formula is too long to be displayed here in print.
%
%  (formula closed but too long
%
%  $$
%-2/3\,\sqrt [3]{-252\,{{t}}^{3}-192\,{{t}}^{2}-192\,{{t}}^{4}-18\,{{t}}^{5}+80
%\,{{t}}^{6}-18\,{t}+80+18\,\sqrt {-96\,{{t}}^{9}-276\,{{t}}^{4}-96\,{{t}}^{3}-
%276\,{{t}}^{8}+192\,{{t}}^{5}-15\,{{t}}^{10}-48\,{{t}}^{12}-96\,{t}-15\,{{t}}^{2}-
%96\,{{t}}^{11}+192\,{{t}}^{7}-48+678\,{{t}}^{6}}}+3/8\, \left( {\frac {256}{
%9}}\,{{t}}^{2}-{\frac {448}{9}}\,{{t}}^{4}-{\frac {64}{3}}\,{t}-{\frac {64}{
%3}}\,{{t}}^{3}-{\frac {448}{9}} \right) {\frac {1}{\sqrt [3]{-252\,{{t}}^{
%3}-192\,{{t}}^{2}-192\,{{t}}^{4}-18\,{{t}}^{5}+80\,{{t}}^{6}-18\,{t}+80+18\,
%\sqrt {-96\,{{t}}^{9}-276\,{{t}}^{4}-96\,{{t}}^{3}-276\,{{t}}^{8}+192\,{{t}}^{5}
%-15\,{{t}}^{10}-48\,{{t}}^{12}-96\,{t}-15\,{{t}}^{2}-96\,{{t}}^{11}+192\,{{t}}^{7}
%-48+678\,{{t}}^{6}}}}}-2\,{t}+{\frac {35}{3}}+{\frac {35}{3}}\,{{t}}^{2}-1/2
%\,i\sqrt {3} \left( 4/3\,\sqrt [3]{-252\,{{t}}^{3}-192\,{{t}}^{2}-192\,{{t}}
%^{4}-18\,{{t}}^{5}+80\,{{t}}^{6}-18\,{t}+80+18\,\sqrt {-96\,{{t}}^{9}-276\,{{t}}
%^{4}-96\,{{t}}^{3}-276\,{{t}}^{8}+192\,{{t}}^{5}-15\,{{t}}^{10}-48\,{{t}}^{12}-
%96\,{t}-15\,{{t}}^{2}-96\,{{t}}^{11}+192\,{{t}}^{7}-48+678\,{{t}}^{6}}}+3/4\,
% \left( {\frac {256}{9}}\,{{t}}^{2}-{\frac {448}{9}}\,{{t}}^{4}-{\frac {64
%}{3}}\,{t}-{\frac {64}{3}}\,{{t}}^{3}-{\frac {448}{9}} \right) {\frac {1}{
%\sqrt [3]{-252\,{{t}}^{3}-192\,{{t}}^{2}-192\,{{t}}^{4}-18\,{{t}}^{5}+80\,{{t}}^
%{6}-18\,{t}+80+18\,\sqrt {-96\,{{t}}^{9}-276\,{{t}}^{4}-96\,{{t}}^{3}-276\,{{t}}
%^{8}+192\,{{t}}^{5}-15\,{{t}}^{10}-48\,{{t}}^{12}-96\,{t}-15\,{{t}}^{2}-96\,{{t}}^
%{11}+192\,{{t}}^{7}-48+678\,{{t}}^{6}}}}} \right)
%  $$
%
Fortunately, the graphical form of the ultimate set of $N=6$
singular values is still easily obtainable. Moreover, it is important
that the form of their time-dependence
(see Figure~\ref{s55pic}) remains
perfectly analogous to its $N=4$ predecessor.

We have to add that the three
branches of the real spectrum of ${X}(t)$ at $t \in (-1,1)$
as well as the absolute values of Im $x_n(t)$ at $t \notin (-1,1)$
are still found localized, asymptotically, inside the three opening wedges
as formed by the singular values
in Figure \ref{s55pic}. The phenomenon of the
bracketing is confirmed.

%is displayed in Figure \ref{555pic}

%
%
%\begin{figure}[t]                     %instead of \begin{figure}[t]
%\begin{center}                         %instead of \begin{center}
%\epsfig{file=zina.eps,angle=270,width=0.6\textwidth}
%\end{center}                         %instead of \end{center}
%\vspace{-2mm}\caption{The $N=6$ sample of the asymptotic spectral
%bracketing as provided by the singular values
%in the case of the complex symmetric
%six by six matrix~(\ref{bufor}).
% \label{555pic}}
%\end{figure}

\subsection{Eight-by-eight matrix $X(t)$\label{8ubtriplsec}}

\begin{figure}[t]                     %instead of \begin{figure}[t]
\begin{center}                         %instead of \begin{center}
\epsfig{file=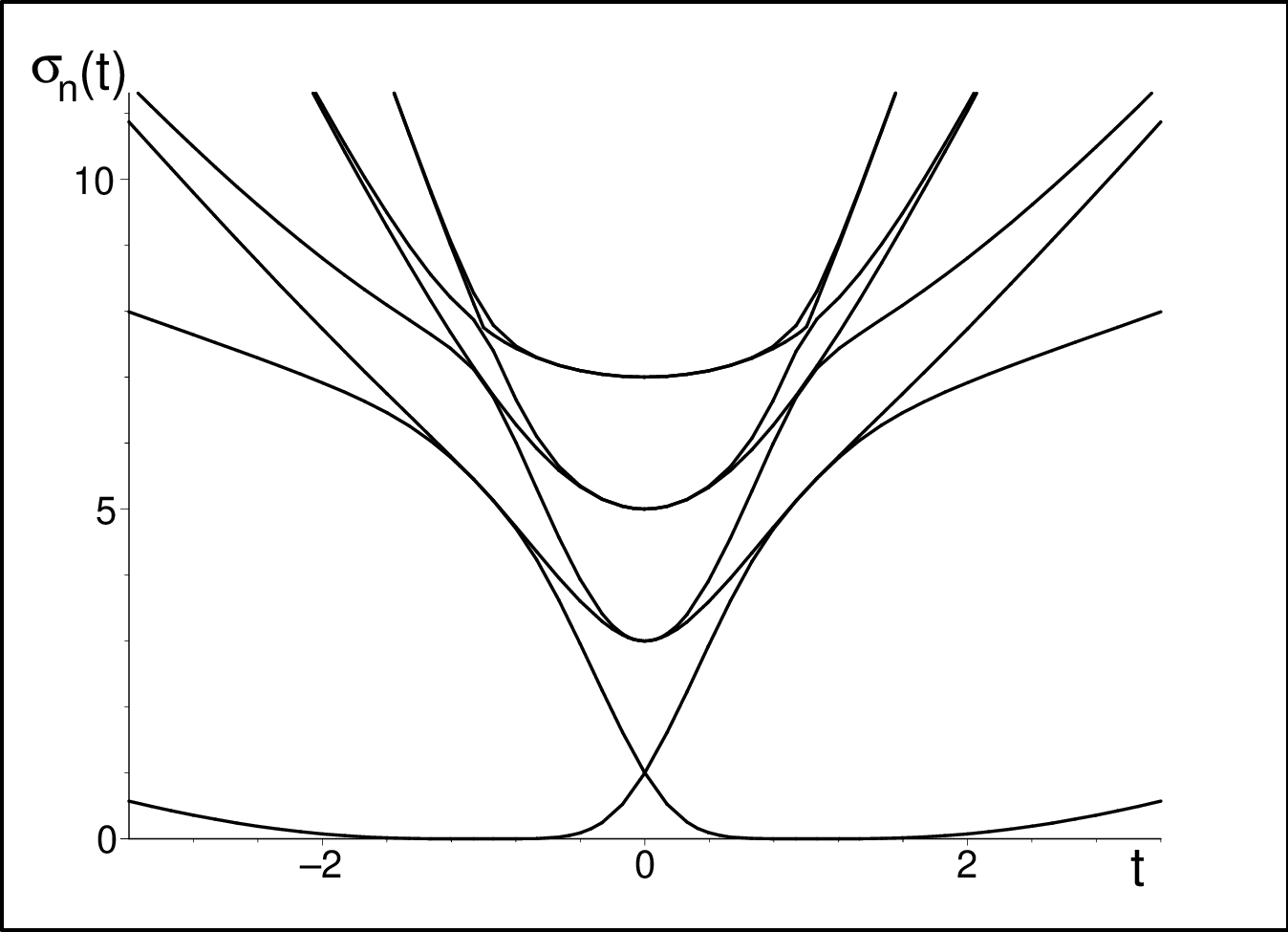,angle=270,width=0.6\textwidth}
\end{center}                         %instead of \end{center}
\vspace{-2mm}\caption{Singular values
for the complex symmetric
eight by eight matrix~(\ref{8bufor}).
 \label{ma55pic}}
\end{figure}

At any $N<\infty$, the
elementary
and exactly known
(i.e., either purely real or purely imaginary)
elements of the spectrum of ${X}(t)$ are made
mutually different just by a multiplication factor \cite{passage}.
It would be, therefore, easy to display their time-dependence
and, if necessary,
to insert these spectral curves
into the above-displayed
graphs of the singular values $\sigma_n(t)$ at $N=4$ and $N=6$.
Nevertheless, once we move to the next, $N=8$ model
 \be
 {X}(t)= \left[ \begin {array}{cccccccc} -7\,i{t}&
 \sqrt {7}&0&0&0&0&0&0\\\noalign{\medskip}\sqrt
 {7}&-5\,i{t}&2\,\sqrt {3}&0&0&0&0&0
\\\noalign{\medskip}0&2\,\sqrt {3}&-3\,i{t}&\sqrt {15}&0&0&0&0
\\\noalign{\medskip}0&0&\sqrt {15}&-i{t}&4&0&0&0\\\noalign{\medskip}0&0&0
&4&i{t}&\sqrt {15}&0&0\\\noalign{\medskip}0&0&0&0&\sqrt {15}&3\,i{t}&2\,
\sqrt {3}&0\\\noalign{\medskip}0&0&0&0&0&2\,\sqrt {3}&5\,i{t}&\sqrt {7}
\\\noalign{\medskip}0&0&0&0&0&0&\sqrt {7}&7\,i{t}\end {array} \right]
 \label{8bufor}
 \ee
one can still feel pleased by the
survival of the existence of the
closed-form
diagonalizability
of the
product
 $
 \mathbb{{X}}={X}^*\,{X}=$
 $$
  \left[ \begin {array}{cccccccc}
  49\,{{t}}^{2}+7&2\,i{t}\sqrt {7}&2\,\sqrt {7}\sqrt {3}&0&0&0&0&0\\
  \noalign{\medskip}-2\,i\sqrt {7}{t}&19+25
 \,{{t}}^{2}&4\,i{t}\sqrt {3}&2\,\sqrt {3}\sqrt {15}&0&0&0&0
 \\\noalign{\medskip}2\,\sqrt {7}\sqrt {3}&-4\,i\sqrt {3}{t}&27+9\,{{t}}^{2
 }&2\,i{t}\sqrt {15}&4\,\sqrt {15}&0&0&0\\\noalign{\medskip}0&2\,\sqrt {3
 }\sqrt {15}&-2\,i\sqrt {15}{t}&31+{{t}}^{2}&8\,i{t}&4\,\sqrt {15}&0&0
 \\\noalign{\medskip}0&0&4\,\sqrt {15}&-8\,i{t}&31+{{t}}^{2}&2\,i{t}\sqrt {15
 }&2\,\sqrt {3}\sqrt {15}&0\\\noalign{\medskip}0&0&0&4\,\sqrt {15}&-2\,
 i\sqrt {15}{t}&27+9\,{{t}}^{2}&4\,i{t}\sqrt {3}&2\,\sqrt {7}\sqrt {3}
 \\\noalign{\medskip}0&0&0&0&2\,\sqrt {3}\sqrt {15}&-4\,i\sqrt {3}{t}&19+
 25\,{{t}}^{2}&2\,i{t}\sqrt {7}\\\noalign{\medskip}0&0&0&0&0&2\,\sqrt {7}
 \sqrt {3}&-2\,i\sqrt {7}{t}&49\,{{t}}^{2}+7\end {array} \right]
 $$
yielding the eight singular values
in the form which is easily stored in the
computer.
%
%
%
%%$$
%%\mathfrak{h}_{11}=5\,\sqrt {1-s^2}=-\mathfrak{h}_{66}\,,\ \ \
%%\mathfrak{h}_{22}=3\,\sqrt {1-s^2}=-\mathfrak{h}_{55}\,,\ \ \
%%\mathfrak{h}_{33}=\sqrt {1-s^2}=-\mathfrak{h}_{44}\,
%%$$
%
%calculated spectrum = explicit eight-level realization of
%the preceding model
%
This means that
the formula for the singular values remains closed
and only too long for a printed display.
In other words, the time-dependence
of these characterstics of the $N=8$ system
%
%
%  $$
%21-2\,s+21\,{s}^{2}+2/3\,\sqrt {6}\sqrt {{\frac {42\,\sqrt [3]{-67544
%\,{s}^{9}+17496\,{s}^{12}+10368\,{s}^{11}-48276\,{s}^{10}+45528\,{s}^{
%7}+12072\,{s}^{4}-67544\,{s}^{3}+12072\,{s}^{8}+10368\,s-48276\,{s}^{2
%}+46888\,{s}^{6}+45528\,{s}^{5}+17496+540\,\sqrt {-432-69732\,{s}^{9}-
%23202\,{s}^{12}+46692\,{s}^{11}+ \ldots}}}{\ldots}}} \ldots
%  $$
%
can still be given, very easily, the
graphical form as sampled by Figure \ref{ma55pic}.
Confirming the trends which characterize
the present choice of the complex symmetric
toy model.

Now, what remains for us to do is
to perform an analogous calculations
for another class of
grid-point-operator matrices.

\section{Another class of models: asymmetric real matrices\label{btriplsec}}

The main material for our independent test
of hypotheses concerning the
Big-Bang-related
singular values
has been found in
our older paper \cite{tridiagonal}
in which we managed to illustrate some of the
benefits as provided by pseudo-Hermitian quantum theory.
For this illustration we proposed and
used certain real asymmetric matrices
which appeared exactly solvable at any matrix dimension $N$.

Let us now
show that these matrices
could also serve for an illustration
of merits of the use of singular values, still
with a particular emphasis upon their applicability and
applications in quantum cosmology.

\subsection{Two by two $X(t)$}

For the purposes of our present
``teaching by example'',
it seems to make sense to recall the
real-matrix
grid-point-operator
 \be
 {X}(t)=\left[ \begin {array}{cc} -1&\sqrt {1-t}\\
 \noalign{\medskip}-\sqrt {1-t}&1\end {array} \right]
 \label{bufo}
 \ee
yielding a very natural
diagonal-matrix equivalent (i.e., isospectral)
avatar
$$
\mathfrak{\xi}(t)= \left[ \begin {array}{cc} \sqrt {t}&0\\
\noalign{\medskip}0&-\sqrt {t}
\end {array} \right]\,
$$
and possessing a unique exceptional point in the origin,
$t^{(EP)}=0$.

\begin{figure}[t]                     %instead of \begin{figure}[t]
\begin{center}                         %instead of \begin{center}
\epsfig{file=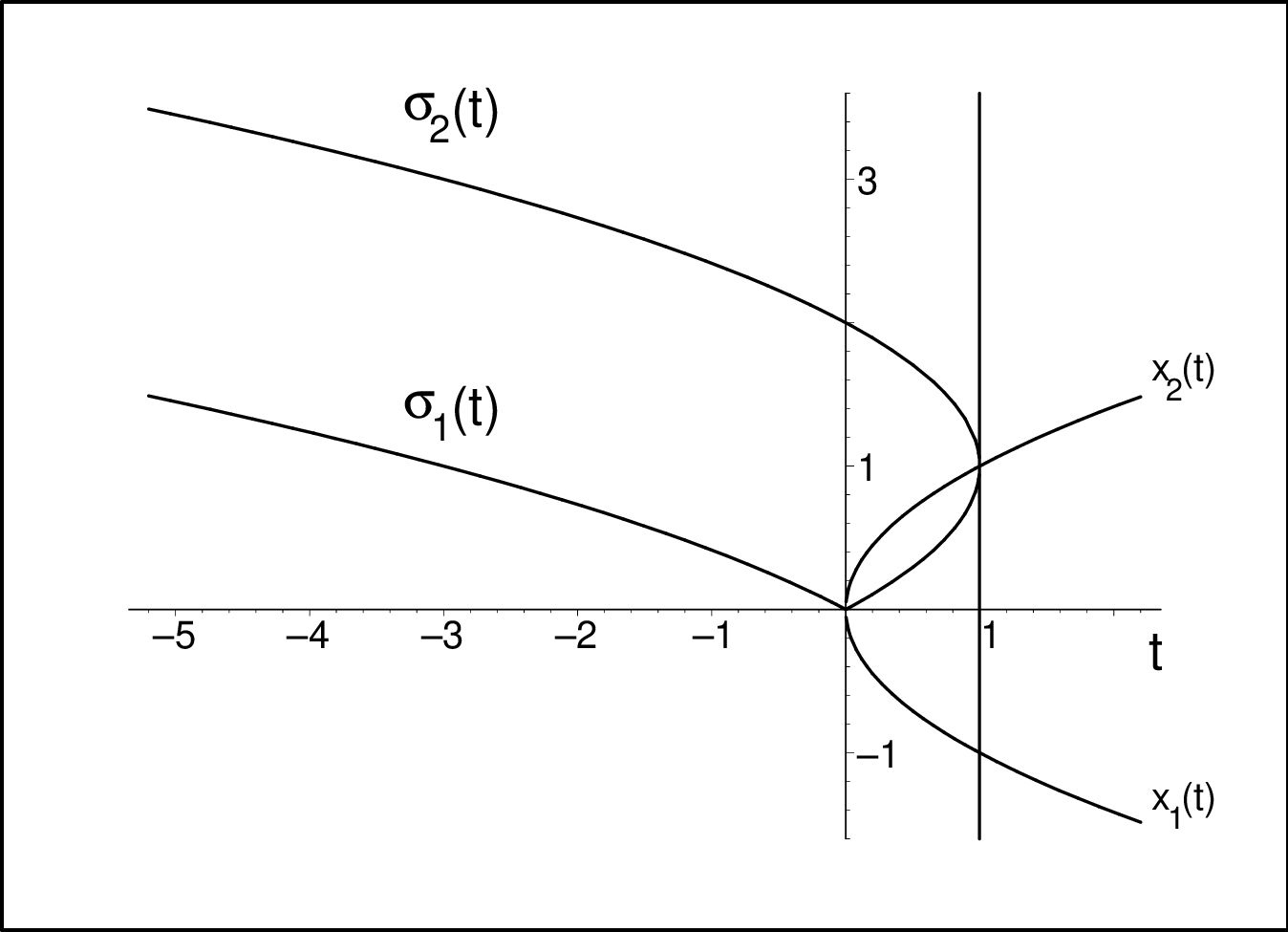,angle=270,width=0.6\textwidth}
\end{center}                         %instead of \end{center}
\vspace{-2mm}\caption{A comparison of the time-dependence of the
real
eigenvalues $x_n(t)$ (= the right-looking parabola)
with the time-dependence of the singular values $\sigma_n(s) $
(= two sections of the two left-looking parabolas)
for the asymmetric but real
matrix
(\ref{bufo}).
 \label{u6pic}}
\end{figure}

Besides the welcome availability of the closed-form
spectrum,
it is also
easy to form the product
 $$
 \mathbb{{X}}(t)={X}^*(t)\,{X}(t)=\left[ \begin {array}{cc} 2-t&-2\,\sqrt {1-t}\\
 \noalign{\medskip}-2\,\sqrt {1-t}&2-t\end {array} \right]
 $$
and to diagonalize it,
  $$
  \widetilde{\mathbb{{X}}}(t)
  =\left[ \begin {array}{cc} 2-t+2\,\sqrt {1-t}&0\\
  \noalign{\medskip}0&2-t-2\,\sqrt {1-t}\end {array} \right]\,.
  $$
Figure \ref{u6pic}  now offers a
comparison
of the two real eigenvalues of ${X}(t)$
(existing at $t \geq 0$ and represented
by the single right-oriented parabola)
with the
two singular values of ${X}(t)$
(represented, at $t<1$, by
the two positive branches of
the two left-oriented parabolas).

In the picture we added a vertical line at $t=1$.
It guides our eye
to the innocent-looking
time $t=1$
beyond which our matrix (\ref{bufo}) becomes Hermitian.
Although such a
``far from Big Bang''
time is not too
interesting for physics (the model is too elementary),
it can serve as an illustration
of the existence of a point of
a change of the overall mathematical paradigm.
Beyond this point, indeed, the
assumption of the non-Hermiticity is broken so that,
as a consequence,
the two singular values degenerate and coincide
with one of the eigenvalues,
viz., with the positive one.
Thus, at $t\geq 1$ one could say that the ``bracketing error''
is zero. Its specification using
singular values becomes
redundant but formally exact.
In such a Hermitian-matrix regime with $t \geq 1$,
therefore,
practically the same information about the system
is carried by the
eigenvalues and by the singular values.

Inside the interval of $t \in (t^{(EP)}_{(BB)},1)$
(with $t^{(EP)}_{(BB)}=0$)
the bracketing
concerns the real eigenvalues of
the non-Hermitian matrix and it
has the same form as above.

%
%\subsection{Big Bang in two-grid-point solvable models\label{atriplsec}}
%
%illustration

% SV - AO

\begin{figure}[t]                     %instead of \begin{figure}[t]
\begin{center}                         %instead of \begin{center}
\epsfig{file=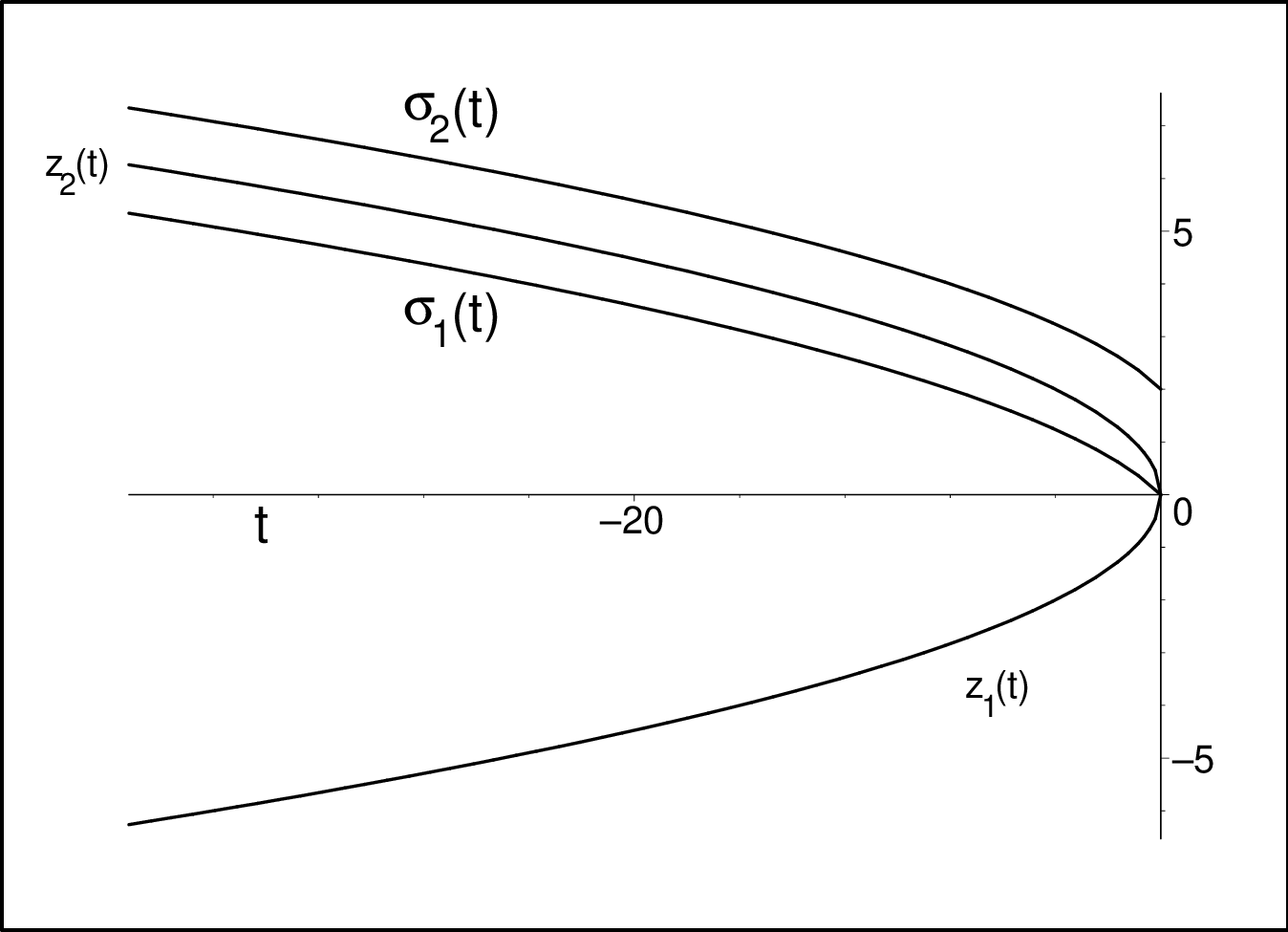,angle=270,width=0.6\textwidth}
\end{center}                         %instead of \end{center}
\vspace{-2mm}\caption{Bracketing
of the purely imaginary eigenvalues
(i.e.,
of
the positive branch of $z_{n}(t)={\rm Im}\,x_n(t)$)
for the real and asymmetric two-by-two
matrix $X(t)$ of Eq.~(\ref{bufo}).
 \label{6pic}}
\end{figure}

What is more interesting is the
confirmation of the expected bracketing
of the absolute value of the complex eigenvalues
at $t<0$ (i.e., before Big Bang).
This can be found illustrated in Figure \ref{6pic}.
In this regime,
the pair of the singular values
offers an upper and lower bound
of the absolute value of the (incidentally, purely imaginary) eigenvalues.

% SV - AO
%
%plot({0.1*sqrt(0+0.1*s^1.99),-0.1*sqrt(0+0.1*s^1.99),0.2*sqrt(0+0.2*s^1.99),
% -0.2*sqrt(0+0.2*s^1.99)},s=-6..12,color=black,axes=framed,tickmarks=[2,2]);
%
%%
%

\subsection{Four by four $X(t)$}

The above-mentioned
toy-model matrix observables are all real and non-Hermitian but still
exactly diagonalizable.
For an illustration of their non-numerical
algebraic tractability
let us now start from their four by four
grid-point-operator
special case
 \be
 {X}({t})=\left[ \begin {array}{cccc} -3&\sqrt {3-3\,{t}}&0&0\\
 \noalign{\medskip}-\sqrt {3-3\,{t}}&-1&2\,\sqrt {1-{t}}&0\\
 \noalign{\medskip}0&-2\,\sqrt {1-
{t}}&1&\sqrt {3-3\,{t}}\\\noalign{\medskip}0&0&-\sqrt {3-3\,{t}}&3
\end {array} \right]
 \label{bufon}
 \ee
with the closed-form Big-Bang-simulating spectrum
 $$
 \{
 -3\,\sqrt {{t}},
 -\sqrt {{t}},
 \sqrt {{t}},3\,\sqrt {{t}}\}
 $$
forming an explicit four-level realization of
its generic sample as displayed in Figure \ref{3pic}.

\begin{figure}[t]                     %instead of \begin{figure}[t]
\begin{center}                         %instead of \begin{center}
\epsfig{file=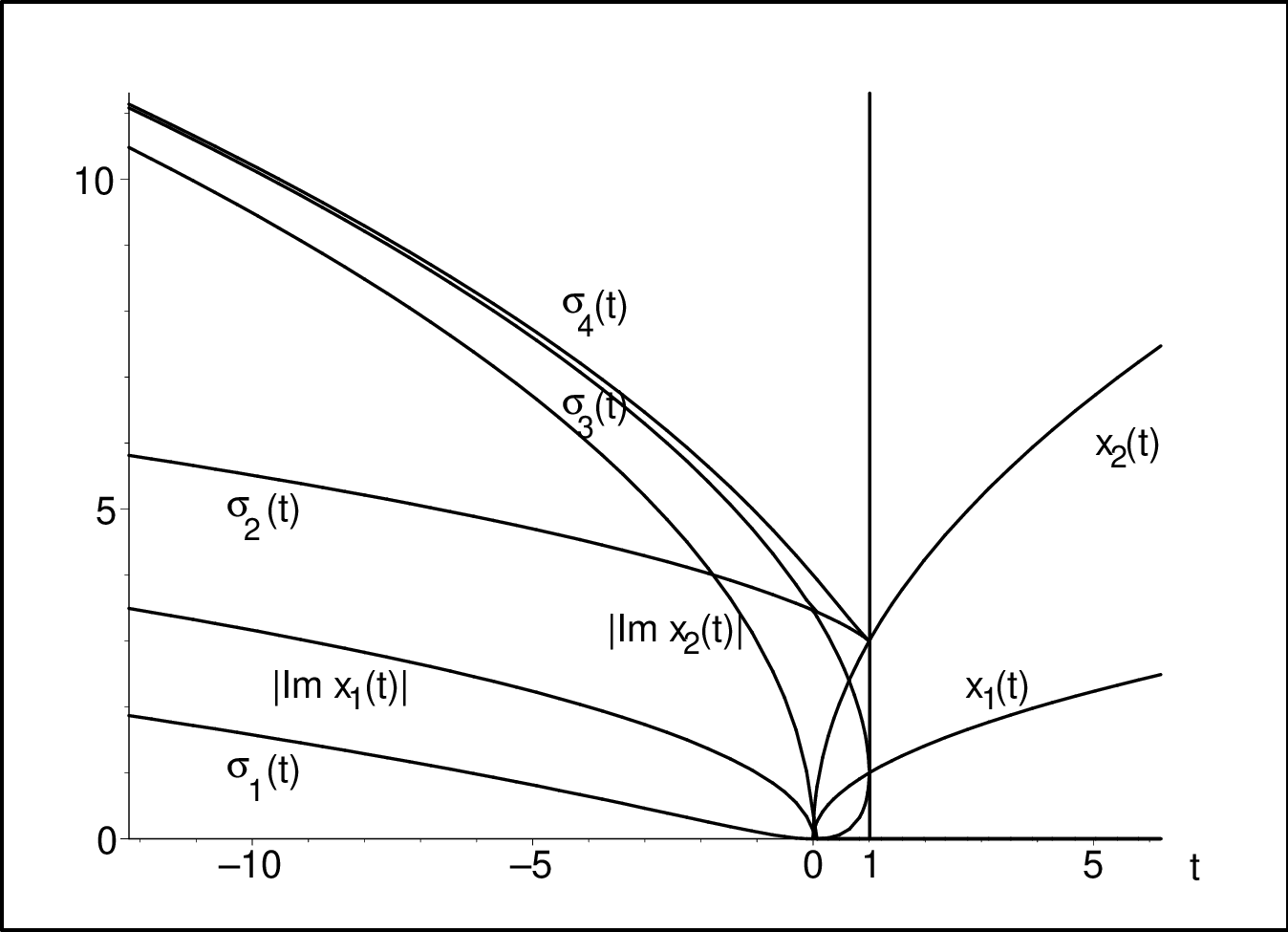,angle=270,width=0.6\textwidth}
\end{center}                         %instead of \end{center}
\vspace{-2mm}\caption{Bracketing relations
for the real and asymmetric
four-by-four matrix~(\ref{bufon}).
 \label{44pic}}
\end{figure}

In a parallel with our preceding four by four
complex symmetric example,
it is still comparatively
easy to form (though not so easy to display)
the product $\mathbb{{X}}(t)={X}^*(t)\,{X}(t)=$
 $$
 =\left[ \begin {array}{cccc} 12-3\,{t}&-2\,\sqrt {3-3\,{t}}&-2\,\sqrt
 {3-3\,{t}}\sqrt {1-{t}}&0\\\noalign{\medskip}-2\,\sqrt {3-3\,{t}}&8-7\,{t}&-4\,
 \sqrt {1-{t}}&-2\,\sqrt {3-3\,{t}}\sqrt {1-{t}}\\\noalign{\medskip}-2\,
 \sqrt {3-3\,{t}}\sqrt {1-{t}}&-4\,\sqrt {1-{t}}&8-7\,{t}&-2\,\sqrt {3-3\,{t}}
 \\\noalign{\medskip}0&-2\,\sqrt {3-3\,{t}}\sqrt {1-{t}}&-2\,\sqrt {3-3\,{t}}
 &12-3\,{t}\end {array} \right]
 $$
yielding the following four closed-form eigenvalues after
non-numerical diagonalization,
  $$
  \sigma_{\pm,\pm}({t})=
  10-5\,{t}\pm 2\,\sqrt {1-{t}} \pm 4\,\sqrt {{\frac {2\,\sqrt {1-{t}}
  -2\,{t}\sqrt {1-{t}}+{{t}}^{2}\sqrt {1-{t}}-2+3\,{t}-{{t}}^{2}}{\sqrt {1-{t}}}}}\ .
  $$
In spite of its closed form, the latter result is much better presented
by the
picture:
In our last Figure \ref{44pic}
the attentive readers may find
not only the singular values but also, for comparison,
both of the
standard eigenvalues $x_n(t)$
which are real for $t>0$ and purely imaginary for $t<0$.

The main point of the message is that
at $t<0$ (i.e., before Big Bang),
the picture offers a very nice illustration of the
inequalities
which represent again the bracketing relations
between the singular values and eigenvalues.

\subsection{Six by six $X(t)$\label{6ubtriplsec}}

%opravit:
%$$
%\mathfrak{h}_{11}=\sqrt {s}=-\mathfrak{h}_{44}\,,\ \ \
%\mathfrak{h}_{22}=3\,\sqrt {s}=-\mathfrak{h}_{33}\,
%$$

The real and asymmetric six by six grid-point-operator matrix
 \be
 {X}=\left[ \begin {array}{cccccc} -5&\sqrt {5-5\,{t}}&0&0&0&0\\
 \noalign{\medskip}-\sqrt {5-5\,{t}}&-3&2\,\sqrt {2-2\,{t}}&0&0&0
\\\noalign{\medskip}0&-2\,\sqrt {2-2\,{t}}&-1&3\,\sqrt {1-{t}}&0&0
\\\noalign{\medskip}0&0&-3\,\sqrt {1-{t}}&1&2\,\sqrt {2-2\,{t}}&0
\\\noalign{\medskip}0&0&0&-2\,\sqrt {2-2\,{t}}&3&\sqrt {5-5\,{t}}
\\\noalign{\medskip}0&0&0&0&-\sqrt {5-5\,{t}}&5\end {array} \right]
 \label{bufec}
 \ee
is to be recalled now showing
the emergence of
the loss of non-numerical solvability.
Although the model remains
non-numerically
solvable for eigenvalues (and the same remains true
at any matrix dimension $N$), the choice of $N=6$
appears to be a boundary
beyond which
the evaluation
of singular values acquires a more or less
numerical character.

A purely formal difficulty even starts when we reveal that  in print,
the product $\mathbb{{X}}={X}^*\,{X}$ appears too large to
fit the standard page.
We only succeed
when we recall the symmetry of this pentadiagonal matrix
with respect to both of its diagonals,
and display just its abbreviated form
%
% $$
% \left[ \begin {array}{cccccc}
% 30-5\,{t}&-2\,\sqrt {5-5\,{t}}&-2\,\sqrt {5-5\,{t}}\sqrt {2-2\,{t}}&0&0&0\\
% \noalign{\medskip}-2\,\sqrt {5-5\,{t}}&22-13
%\,{t}&-4\,\sqrt {2-2\,{t}}&-6\,\sqrt {2-2\,{t}}\sqrt {1-{t}}&0&0
%\\\noalign{\medskip}-2\,\sqrt {5-5\,{t}}\sqrt {2-2\,{t}}&-4\,\sqrt {2-2\,{t}
%}&18-17\,{t}&-6\,\sqrt {1-{t}}&-6\,\sqrt {2-2\,{t}}\sqrt {1-{t}}&0
%\\\noalign{\medskip}0&-6\,\sqrt {2-2\,{t}}\sqrt {1-{t}}&-6\,\sqrt {1-{t}}&18
%-17\,{t}&-4\,\sqrt {2-2\,{t}}&-2\,\sqrt {5-5\,{t}}\sqrt {2-2\,{t}}
%\\\noalign{\medskip}0&0&-6\,\sqrt {2-2\,{t}}\sqrt {1-{t}}&-4\,\sqrt {2-2\,
%{t}}&22-13\,{t}&-2\,\sqrt {5-5\,{t}}\\\noalign{\medskip}0&0&0&-2\,\sqrt {5-5
%\,{t}}\sqrt {2-2\,{t}}&-2\,\sqrt {5-5\,{t}}&30-5\,{t}\end {array} \right]
% $$
%
%alias
 $$
 \left[ \begin {array}{ccccc}
 30-5\,{t}&-2\,\sqrt {5-5\,{t}}&2\, \left( -1+{t} \right) \sqrt {5}\sqrt {2}
 &0&\ldots
 %&0&0
 \\
 \noalign{\medskip}-2\,\sqrt {5-5
 \,{t}}&22-13\,{t}&-4\,\sqrt {2-2\,{t}}&6\, \left( -1+{t} \right) \sqrt {2}&\ddots
 %&0&0
 \\
 \noalign{\medskip}2\, \left( -1+{t} \right) \sqrt {5}\sqrt {2}&-4\,
 \sqrt {2-2\,{t}}&18-17\,{t}&-6\,\sqrt {1-{t}}&\ddots
 %&6\, \left( -1+{t} \right) \sqrt{2}&0
 \\
 \noalign{\medskip}0&6\, \left( -1+{t} \right) \sqrt {2}&-6\,
 \sqrt {1-{t}}&18-17\,{t}&\ddots
 %&-4\,\sqrt {2-2\,{t}}&2\, \left( -1+{t} \right) \sqrt {5}\sqrt {2}
 \\
 \noalign{\medskip}0&0&6\, \left( -1+{t} \right) \sqrt {2}&
 -4\,\sqrt {2-2\,{t}}&\ddots
 %&22-13\,{t}&-2\,\sqrt {5-5\,{t}}
 \\
 \noalign{\medskip}0&0&0
 &2\, \left( -1+{t} \right) \sqrt {5}\sqrt {2}&\ldots
 %&-2\,\sqrt {5-5\,{t}}&30-5\,{t}
 \end {array} \right]\,.
 $$
Using the computer-assisted algebraic manipulations we can still
deduce a printable secular polynomial
 $$
  P(x,{t})=
 {x}^{6}+ \left( 70\,{t}-140 \right) {x}^{5}+
 \left( 1743\,{{t}}^{2}-
 7728\,{t}+7728 \right) {x}^{4}+
 $$
 $$
 + \left( 18580\,{{t}}^{3}-142152\,{{t}}^{2}+314976\,{t}-209984 \right) {x}^{3}+
 $$
 $$
 + \left(82831\,{{t}}^{4} -998960\,{{t}}^{3}+3804720\,{{t}}^{2}
 -5611520\,{t}+2805760
 \right) {x}^{2}+
 $$
 $$
 +\left(116550\,{{t}}^{5}-2191500\,{{t}}^{4}+13248000\,{{t}}^{3}
   -33408000\,{{t}}^{2}+36864000\,{t}-14745600
 \right) x+50625\,{{t}}^{6}
 $$
but its roots (i.e., the sextuplet of the real
and time-dependent eigenvalues of $X(t)$)
can only be localized numerically.

We may conclude that
in contrast to its complex predecessor,
the alternative real and asymmetric toy model ceases to be
suitable for our present methodical purposes at $N \geq 6$.

\section{Discussion\label{discussion}}

\subsection{Classical versus quantum
gravity in cosmology\label{uvprecese}}

The current state of relationship
between the classical and quantum gravity
can briefly be characterized
by the Thiemann's words that
``despite an enormous effort of work
by a vast amount of physicists \ldots
we still do not have a credible quantum
general relativity theory''
because ``the problem is so hard'' \cite{Thiemann}.
In this area of research, therefore, one has to
appreciate even a partial progress.
An amendment of the insight
in the multiple open
problems
connected with the quantization of gravity
can really be achieved via the detailed technical
analyses of certain suitable simplified models.

A few results inspired by these questions were also presented in our
present paper. In fact, the background of our considerations
can be traced back to Figure \ref{1pic}
indicating that
after one makes
the most conventional choice of metric $\Theta=\Theta_0=I$
(which means that all of the related
admissible operators of observables
must be Hermitian,  $\Lambda_0=\Lambda_0^\dagger$ \cite{Geyer}),
the existence of any
Big-Bang-like degeneracy
would be excluded,

Whenever $\Theta=I$,
this
is a no-go statement which
holds unless we impose a suitable
{\it ad hoc\,} symmetry. Thus,
the typical parameter-dependence of eigenvalues
would have the form of avoided crossing.
Mathematicians would say that
due to the Hermiticity of the operator,
the related Kato's
EP-localizing parameter
(i.e., in our models,  the time of Big Bang)
must necessarily be complex.

In contrast, after one relaxes the Hermiticity constraint,
one finds (or can construct) multiple examples in which
$\Theta \neq I$ is nontrivial,
and in which the
critical
EP parameter becomes real.
Then, by definition the level-crossing may become
exact, unavoided.

One of the scenarios of the later type
has constructively been studied, e.g., in \cite{passage}.
We used there, exclusively, the very special
parametric dependence of
the EP-supporting
spectra as sampled here in Figure \ref{2pic}.
By construction, these spectra were guaranteed real
both before and after the passage through the EP singularity.
In the picture
one can see a realization of the
Penrose's \cite{Penrose} eternally-cyclic evolution.
By this hypothesis,
there exists a sequence of well-separated \textcolor{black}{Aeons}
of evolution, every one initiated by the
Big Bang re-birth of the Universe.

In order to make all of these considerations less speculative
and, at the same time, feasible,
we turned attention
from the spectra
to the operators.
Emphasis was put on the study of
their phenomenological and mathematical aspects.
In particular,
the conventional
construction of eigenvalues
was complemented
by a less conventional
study of singular values.
We showed that even the reconstruction
of an incomplete
information carried by
the latter quantities enables one to
characterize the dynamics of
the system even
in the anomalous regime
near Big Bang
in which
the observability may get lost.

We managed to show that
even after the
loss of the reality of
at least some of the eigenvalues,
the
singular values
themselves remain, by definition, real.
From a purely pragmatic point of view, this
was shown to open the way towards
a reinterpretation of physics of quantum cosmology
in terms of inequalities.

\subsection{Quantum Big Bang in schematic picture\label{jednasec}}

Due to an enormous complexity of the gravity-quantization challenge,
one can notice a sharp contrast between the
abstract projects and explicit predictions.
One of the decisive specific conceptual
obstacles emerging in the
quantum theory of gravity near Big Bang
is the necessity of having this theory
``background independent''.
This means that in contrast to the relativistic
quantum electrodynamics
or quantum chromodynamics,
one cannot
assume the existence of a fixed space-time framework
(and
perform the quantization
of a time- and coordinate-dependent field $\phi(\vec{x},t)$)
because the geometry of the space-time becomes, by itself,
field- and time-dependent \cite{Thiemann}.

One of the ways out of the dead end has been found in
the so called loop quantum gravity \cite{Ashtekar}.
A
consequent background independence is achieved there
at the expense of an enormous increase of the
complexity of the formalism. Thus, not too surprisingly,
even the loop-quantum-gravity-based answers to
the rather fundamental
question of the existence or non-existence of an
initial Big Bang singularity
remain ambiguous \cite{BBzpet,piech}.

%The overall
%conceptual satisfaction evoked by the success of  the loop quantum gravity
%is still accompanied by the
%absence of ultimate answers.
In such a broader conceptual framework
the essence of the innovation
as provided by the present approach
is that under certain appropriate mathematical conditions,
some of the
observables
(say, $\Lambda$) may be assigned a physical inner-product metric
$\Theta=\Theta(\Lambda)$.

Both $\Lambda$ and $\Theta$ operators can
be kept parameter-dependent
and, in particular, time-dependent.
The formalism seems particularly suitable for
the analysis of one of the
most important quantum-cosmology problems, viz.,
of the question of compatibility between the classical and quantum
notion of Big Bang.

\subsection{Wheeler-DeWitt equation\label{dnasec}}

Among the numerous existing
steps of partial progress
we felt
inspired by the observation that
in the particular context of quantum cosmology
one of the key points remains to be
``the problem of finding an appropriate inner product
on the space of solutions of the
Wheeler-DeWitt equation'' \cite{ali}.

The
reconstructions of
phenomenologically acceptable ``physical'' inner products
for the
Wheeler-DeWitt fields
offered a deeper
new insight in the possible
consistent probabilistic interpretation of the
canonical gravity
in the
context of quantum theory in its pseudo-Hermitian-operator
representation -- cf., e.g., \cite{ali}.
The author of the latter review paper
also added that
``most of the practical and conceptual difficulties
of addressing the \ldots problems for \ldots Wheeler-DeWitt fields
can be reduced to, and dealt with, in the context of [a certain] simple
oscillator'' (cf. section Nr. 3.5 or page 1293 in \cite{ali}).

Needless to add that
such a
``simple
oscillator'' was just a schematic,
methodically oriented model,
not designed to mimic any
empirically relevant phenomena.
Still,
the conceptually innovative
pseudo-Hermitian-operator nature of the
underlying
Wheeler's and DeWitt's ``pseudo-Hermitian
Hamiltonians'' $H \neq H^\dagger$
has to be considered encouraging.

In this sense, our present paper can be also read as an immediate continuation
of the
two-by-two matrix example
as given in equation Nr. 379 of review~\cite{ali}.
\textcolor{black}{An even more persuasive support of the
potential relevance of the present grid-point philosophy
could be sought, in {\it loc. cit.},
in the preceding ``generic'' equation Nr. 377
which is often considered in the partial differential equation
form in applications.
In such a setting, naturally,
one reveals the existence of a certain tension between the
discrete grid-point nature
of our present models
and the continuous-limit essence of the physical reality.
Many questions emerge, especially those concerning the
optimality of the grid-point lattice
near the singularities, etc.}

\subsection{The concept of pseudo-Hermiticity\label{urecese}}

There exists a specific merit of
such an overall theoretical framework
which lies in an innovation of the
form of
information about the observable aspects of the system in question.
This information
is carried
by a pair of operators
(thus, let us speak here about $\Lambda$ and $\Theta$)
in a way
extending the scope of
conventional
textbooks in which such an information is
carried just
by a single dedicated operator $\Lambda$
called ``observable''~\cite{Messiah}.

The eligible metrics $\Theta$
must be such that the operator product $\Theta\,\Lambda$ is
self-adjoint. A more complete but still concise specification of the
necessary mathematics can be found, e.g., in \cite{Geyer}.
In this abstract theoretical framework
one of the two information-carrying operators
remains to be the conventional one (sampled, e.g.,
by the Hamiltonian
$H=\Lambda_0$
representing the observable energy
in Schr\"{o}dinger picture).
Its new, {\it ad hoc\,} partner $\Theta$
(called the physical inner product metric)
can be then selected from a
fairly large set of suitable, $\Lambda-$compatible
candidates.

This is precisely what opens the way towards
a mathematically fully consistent presence of
an EP singularity in $\Lambda$.
In the particular physical quantum-Big-Bang context
the core of relevance of a nontrivial
operator of metric
(which would coincide with the identity operator
in conventional setting)
lies in its variability.
Briefly stated,
any EP-related
degeneracy
(and, in particular, the Big-Bang spatial degeneracy)
can be compensated, in a
properly fine-tuned manner, by a
properly adapted $\Theta=\Theta(\Lambda)$ \cite{SciRep}.

\textcolor{black}{Needless to add that the internal consistency of
such a basic form of the
pseudo-Hermitian quantum theory
opens the way towards an extension
of its applicability and, in particular, towards
the possibility of study, say, of
the thermodynamic properties of the
Big Bang scenario and/or scenarios.
Although these questions already lie beyond the scope of our present paper,
investigations can be expected of the concepts of
entropy and/ofr temperature, especially near the EP.
One can also expect success in the search for connections
of the EP-based theory with
the currently quickly developing thermodynamics
of black holes, etc.}

\subsection{Singular values and simplifications}

In our recent letter \cite{MCF}
devoted to the pseudo-Hermiticity-related numerical methods we proposed
a replacement of the study of the complex eigenvalues of
operators
by the less ambitious search
for their singular values.
We revealed that these quantities
(defined as square roots of eigenvalues of the
self-adjoint products ${X}^\dagger \,{X}$,
i.e., still carrying a significant
amount of relevant information about ${X}$ itself)
are all real, so that
their numerical localization
becomes more easy and user friendly.

Our present paper described a number of
examples
for which the study of singular values remained,
in a way
motivated
by the phenomenology of Big Bang in cosmology, non-numerical.
Sharing the maximally user-friendly
tractability by the pure linear-algebraic means.
In spite of the
simplifications (which are, from the point of view of
any more realistic
consideration,
rather drastic) their
main merit lies in the insight in the
EP-related loss or return of the diagonalizability.

Besides the manifestly non-covariant nature of our toy models,
their
most efficient simplification should be seen in our replacement
of
the realistic 3D
spatial background
by the mere 1D discrete
few-grid-point lattice.
Its elements are treated as mimicking the evolution of
the expanding Universe after
the point-like Big Bang at $t=0$.

Such a representation
of the physical reality does not take into account the dynamics of
any energy-momentum-carrying fields because
we only introduced certain artificial time-dependent
spatial geometry
which is assumed quantized, i.e.,
treated
as a conventional quantum-dynamical observable.
Every element of
the set of spatial grid points
is interpreted
as a time-dependent eigenvalue of a
self-adjoint operator $X(t)$ acting in
Hilbert space ${\cal H}_{}$.
As an immediate consequence,
time $t$ is treated
as a mere real parameter without quantum interpretation.
\textcolor{black}{Naturally,
the time treated
as a mere parameter rather than a
dynamical coordinate does really oversimplify the intricate relationship
between the
space-time geometry and the mathematics inherent in quantum
gravity}
(in this respect see also
  %the footnote ``bbbb'' %on
p. 1292 in \cite{ali}).
The discretization of the
coordinates ${x} \to x_k$, $k \in {\mathbb{N}}$ yielded
the equidistant and finite grid-point lattice.

%\subsection{Questions\label{esse}}
%
%There are two questions which emerge,
%after the methodically motivated simplifications, immediately.
%The first one

\textcolor{black}{Concerning} the
feasibility of reinstallation of
singularity after quantization
%In this respect
we accepted the
philosophy
of the pseudo-Hermitian quantum
mechanics.
We revealed that
such an approach to realistic cosmology
profits
from the enhancement of
flexibility
of the model-building processes.

In fact, the main gain appeared to concern the
preservation of singularities.
In the language of mathematics
the spatial Big Bang degeneracy
 \be
 \lim_{t \to t_{BB}}x_k(t)=x_{BB}, \ \ \ \forall k\,.
 \ee
has been found tractable
as reflecting the exceptional-point value of time.

Another emerging question
concerns the differences
between the
alternative
pre-Big-Bang dynamics.
This problem
was sampled by several pictures,
all of which shared the assumption
of reality of the
eigenvalues of $X(t)$ after Big Bang.
In all of them, still,
the specific alternative choices of
the structure
of the pre-Big-Bang era
can only be perceived as speculative.

In this respect we felt inspired by the
conventional nuclear, atomic or molecular physics
and by the
widely accepted treatment of
differences between the so called closed systems (i.e., stable, bound-state
systems)
and the so called  open systems
(i.e., unstable, resonant
quantum systems).
In the latter cases, indeed, there exist
multiple forms of the comparison of the theory with experiments,
%
%For our present purposes
%of study of the quantum Universe near its Big Bang singularity
%we turned attention, therefore,
\textcolor{black}{with some of them related}
to
the evaluation of the less usual
mathematical characteristics
of the system
like singular values
\cite{Nimrod,Ingrid}.

\section{Summary\label{summary}}

Once one decides to use the Kato's exceptional points
as a means of quantization of classical singularities,
the first task which must be fulfilled is a description and
classification of the ways of their unfolding.
This opens the way
towards our understanding of a
wealth of quantum phase-transition scenarios
ranging from
the condensed-matter physics
up to the cosmology and quantum Big Bang.

A key to our present message lies in the observation that
one of the really efficient ways of a clarification of
an overall mechanism of these unfoldings
is provided by the study of
elementary non-numerical toy models.
Formally, we redirected this study
from the conventional analysis of spectra
(which are, in general, complex and difficult to localize)
to a reduced description of complicated quantum systems
in terms of singular values.

The phenomenological context was
sampled here by
the cosmological
Big Bang.
The assumption of existence of the point-like form of
such a degeneracy has been shown to
open a direct access to a broad
family of new physical quantum-phase-transition-like phenomena.
At the same time, multiple new
questions emerged.

A physical motivation
of their present analysis was threefold.
Firstly,
we were impressed
by paper \cite{aliKG} which
elucidated the
connection
between the innovative pseudo-Hermitian reformulation
of quantum mechanics
and
the canonical quantization of gravity.
This made us believe that
a deeper analysis of solvable models
can offer the answers to
a few relevant methodical as well as conceptual questions.

Secondly,
we recalled the
pseudo-Hermiticity-related possibility of
identification of the
quantum phase transitions
with the EP-type spectral
degeneracies.
Although this trick can mostly be
found in optics rather than in cosmology
\cite{Berry,Mousse,Christodoulides},
we emphasized that
all of the related
technicalities
could directly be extended to
cover the Big-Bang-related
phenomena.
After all, precisely the same transfer of experience
is used in
several other branches of physics ranging from
classical optics
up to the broad area of
atomic, molecular or nuclear
physics \cite{Heiss,Nimrod,Ingrid}.

Thirdly,
we revealed that the simulation of quantized Big Bang
could profit from the knowledge of
the concept of singular value.
During the formulation of our present project
we felt encouraged by the
parallel developments of this topic
in mathematics.
Quick progress could have been seen there not only in
the abstract EP-related spectral theory
of non-Hermitian operators
\cite{PS1,book}
but also
in its various
(i.e., e.g.,
numerical \cite{Nimrod,MCF,Marcelo} or
theoretical \cite{Nimrodd})
implementations.

What remains to be particularly appreciated is that
in the recent literature there appeared a gap between
the existence of studies of the unitary evolution channels
(i.e., of the corridors in the space of parameters
in which the spectrum remains real \cite{passage})
and the absence of analogous model-based
illustrative descriptions of the
EP unfoldings in which the (discrete) spectrum is
more general (i.e., complex).

From several
complementary
points of view
the gap has been partially filled here.
In particular,
what seems to be truly promising for the future research
is the progress achieved,
via simplified models, in our understanding of operators which lie
in a ``small'' vicinity
of
an EP singularity.
In this vicinity (whatever its ``smallness'' could mean \cite{pert}),
the structure of evolution gets split into a broad
variety of patterns yielding, sometimes, very specific EP unfoldings.
Their bracketing-based
classification emerges which works with
singular values and which seems able to
distinguish, even in the quantum-gravity context,
between the conventional
unitary-evolution scenarios
and their resonances-based open-system-like alternatives.

%\section*{Acknowledgement}
%
%The author was supported by
%Faculty of Science of
%University of Hradec Kr\'{a}lov\'{e}

\end{document}